\documentclass{aa}    
\begin{document}
\newcommand{\lesssim}{\stackrel{<}{\sim}}
\newcommand{\gtrsim}{\stackrel{>}{\sim}}
\newcommand{\be}{\begin{equation}} 
\newcommand{\ee}{\end{equation}} 
\newcommand{\der}[2]{\frac{d{#1}}{d{#2}}} 
\newcommand{\dertext}[2]{{d{#1}}/{d{#2}}} 
\newcommand{\pd}[2]{\frac{{\partial}{#1}}{{\partial}{#2}}} 
\newcommand{\pdtext}[2]{{{\partial}{#1}}/{{\partial}{#2}}} 
\newcommand{\func}[1]{\mathrm{#1}} 
\newcommand{\rot}[1]{\mathrm{rot}{#1}} 
\newcommand{\dvrg}[1]{\mathrm{div}{#1}} 
\newcommand{\en}{\varepsilon} 
\newcommand{\nlb}{{\nolinebreak}} 
\newcommand{\kap}{\symbol{"1A}} 
\title{Eigenoscillations of the differentially rotating Sun: I.\\
 22-year, 4000-year, and quasi-biennial modes}
\author{N.S. Dzhalilov\inst{1,}\inst{2} \and J. Staude\inst{2} \and V.N.
Oraevsky\inst{1}} 
\offprints{J.Staude} 
\institute{Institute of Terrestrial Magnetism,
Ionosphere and Radio Wave Propagation of the Russian Academy of Sciences,\\
Troitsk City, Moscow Region, 142190 Russia; 
E--Mail: namig@izmiran.rssi.ru, oraevsky@izmiran.rssi.ru
\and Astrophysikalisches Institut Potsdam, Sonnenobservatorium Einsteinturm,  
14473 Potsdam, Germany;\\
 E--Mail: jstaude@aip.de}
\date{Received\ \ \ \ \ \ \ \ \ ; accepted\ \ \ \ \ \ \ \ }

\abstract{
Retrograde waves with frequencies much lower than the rotation frequency
become trapped in the solar radiative interior. The eigenfunctions of the
compressible, nonadiabatic, Rossby-like modes ($\epsilon$-mechanism and
radiative losses taken into account) are obtained by an asymptotic method
assuming a very small latitudinal gradient of rotation, without an
arbitrary choice of other free parameters. An integral dispersion relation
for the complex  eigenfrequencies is derived as a solution of the boundary
value problem. The discovered resonant cavity modes (called $R$-modes) are
fundamentally different from the known $r$-modes: their frequencies are
functions of the solar interior structure, and the reason for their existence
is not related to geometrical effects. The most unstable $R$-modes are those
with periods of $\approx$\,1--3\,yr, 18--30\,yr, and 1500--20\,000\,yr;
these three separate period ranges are known from solar and geophysical data.
The growing times of those modes which are unstable with respect to the
$\epsilon$-mechanism are $\approx 10^2, 10^3,$ and $10^5$ years,
respectively. The amplitudes of the $R$-modes are growing towards the center
of the Sun. We discuss some prospects to develop the theory of $R$-modes as a
driver of the dynamics in the convective zone which could explain, e.g.,
observed short-term fluctuations of rotation, a control of the solar magnetic
cycle, and abrupt changes of terrestrial climate in the past. %
}

\titlerunning{Long-period eigenoscillations of the Sun. I.} 
\authorrunning{N.S. Dzhalilov, J. Staude \& V.N. Oraevsky} 
\maketitle


   \section{Introduction}
      

The 22-year magnetic cycle of solar activity is the most prominent
phenomenon of several large-scale dynamic events that occur in the Sun.
(Really, the magnetic half cycles or sunspot number cycles vary in length
between 9--13 years, and 11 yr is an average of the $\approx20$ half-cycles
available.) An explanation of the basic mechanism underlying this fameous
phenomenon is the fundamental challenge of solar physics. The achievements of
the theory of the $\alpha$-$\omega$ dynamo turned out to be a great success.
However, neither all observations of magnetic and flow fields nor the
radiation fluxes which are related to this phenomenon and which are measured
at the surface of the Sun or indirectly, by helioseismology, in its interior,
can be explained unambigously in this way. Although our present work is not
directly related to the dynamo theory, we will outline here those
difficulties which have common points with our results.


\subsection{Some problems of dynamo theory }.
      

As a consequence of our imperfect knowledge of basic characteristics of
turbulent convection as well as meridional circulation and details of the
rotation of the Sun's interior, the solutions of the dynamo equations become
functions of many free, unknown parameters (e.g. Stix 1976). For instance, by
clever combinations of these parameters it is possible to get from kinematic
theory an oscillatory magnetic field with a 22-year period and a growing
amplitude. However, another choice of these parameters leads to waves of
growing amplitude for other periods. So one could draw a butterfly
diagram not only with an 11-year periodicity. It remains still an open
question which of the clever combinations resulting in a solar-like 22-year
activity cycle is realized in the Sun. We could not find a work on the dynamo
wave problem, showing that just the 22-year period is preferred among others
with a maximum growth rate and with the spatial scales required for solar
activity. Instead, many authors pointed out that the cycle period of 22 years
is hard to explain (Stix 1991; Gilman 1992; Levy 1992; Schmitt 1993;
Brandenburg 1994; Weiss 1994; R\"udiger \& Arlt 2000). 

From the solution of the inverse problem of helioseismology (e.g. Tomczyk et
al. 1995) it is known that the convective envelope of the Sun is rotating
with a latitude dependence of the angular velocity similar to that of the
surface but almost rigid in radial direction. A stronger radial gradient
which is required for the $\alpha$-$\omega$ dynamo mechanism is located in a
shallow layer (thickness $\approx 0.05 R_{\odot}$ (Kosovichev 1996), where
$R_{\odot}$ is the solar radius) immediately below the convective zone ---
the tachocline (Spiegel \& Zahn 1992). Below the tachocline up to a depth of
at least $0.5 R_{\odot}$ the radiative interior is rotating with an angular
velocity law similar to that of a solid-body. The  question arises: what
compels the Sun to rotate in such a strange manner, which is different from
the generally accepted, theoretically predicted stable rotation law? How to
handle a dynamo theory for which the `$\omega$' area is separated from the
`$\alpha$' area over a large part of the extent of the convective zone? In
order to solve this problem Parker (1993) has put forward the idea of an
interface dynamo, the basic features of which existed already in earlier
dynamo models (Steenbeck et al. 1966). To close the cycle of such a stretched
dynamo it is necessary to have some mechanism delivering toroidal magnetic
flux, arising by the shear of differential axisymmetric rotation (Cowling 1953)
in the tachocline, to the `$\alpha$' dynamo area (e.g. Moffatt 1978; Krause \&
R\"adler 1980). To get a solar-like magnetic activity it is necessary to
suppose the existence of a huge ($\approx 10^5$\,G) toroidal magnetic field
to create enough magnetic buouancy for the leakage of magnetic flux and to
solve the tilt problem of lifting loops (e.g. Caligari et al. 1998).
Moreover, a high magnetic diffusivity contrast between the convective
envelope and the underlying radiative core should be assumed to solve the
quenching problem of the $\alpha$ effect (see, e.g., Fan et al. 1993;
Cattaneo \& Hughes 1996). However, it is a major challenge for any dynamo
model to produce such strong fields.

The idea of the interface dynamo was further developed, e.g. by
Charbonneau \& MacGregor (1997). Later, a fit to the real solar rotation
profile with its latitudinal and radial dependencies has been included by
Markiel \& Thomas (1999), but so far
no satisfactory solar-like oscillatory solutions for the interface dynamo
have been found. Growing wave solutions are suppressed by the
latitudinal shear.  


\subsection{Spinning-down rotation problem}.
      

Mechanisms for braking the solar internal rotation are also under discussion.
The  character of the core rotation is not clear because here the accuracy
of helioseismic inversions gets worse (Chaplin et al. 1999) and the results
seem to be in contradiction with the oblateness measurements (Paterno et al.
1996). There are some suggestions that a deceleration of the radiative
interior depends on the transport of angular momentum between this region and
the convective zone. For instance, Mestel \& Weiss (1987) supposed that even
a weak large-scale magnetic field would be sufficient to couple very
efficiently the interior and the  convective zone, leading essentially to
solid body rotation. In this way the magnetic torques can also extract
angular momentum from the  radiative interior (e.g. Charbonneau \& MacGregor
1993). 

The wave mechanism for the solution of this problem is more popular.
Schatzman (1993), Zahn et al. (1997), and Kumar \& Quataert (1997) have
concluded that the solid rotation of the radiative interior is a direct
consequence of the effect of internal gravity waves. Gravity waves generated
near the interface between the convective and radiative regions transport
retrograde angular momentum into the interior, thereby spinning it down. Here
the main idea is that the isotropically generated gravity waves become
anisotropic due to Doppler shifts of frequencies in the differentially
rotating Sun. In that way for anisotropic retrograde and prograde waves the
radiative damping is different, and the residual negative angular wave
momentum may compel the solar radiative interior to co-rotate with the
convective zone. This idea has been further developed by Kumar et al. (1999)
including a toroidal magnetic field to explain the existence of the unstable
shear layer `tachocline'. However, Ringot (1998) has shown that a quasi-solid
rotation of the radiative interior cannot be a direct consequence of the
action of internal gravity waves produced in the convective zone. Gough
(1997) questioned this idea emphasizing that the mechanism can work only if
the waves are generated with strong amplitudes to transport the required
angular momentum. This means, resonance waves are required, but such waves
may penetrate only to distances less than $10^{-5}R_{\odot}$ beneath the
convective zone due to the strong radiative damping. These waves must deposit
their angular momentum before returning to the convective zone, but not
before penetrating far into the radiative interior.

For the wave mechanism the question of an anisotropic propagation
relative to the azimuthal rotation is a key moment. Fritts et al. (1998) have
shown that convection, penetrating into the stratified and strongly sheared
tachocline, can produce preferentially propagating gravity waves.

There have also been speculations that the rotation of the core may be
variable, perhaps with a time scale of the solar cycle (e.g. Gough 1985).
The present paper is along these lines.

From our short discussion we conclude that the convective envelope and the
radiative interior are coupled to each other through a certain global agent,
resulting in almost co-rotation. To advance the solution of the problem the
dynamo theory should take into account the presence of this global agent. We
suppose that really this agent is provided by waves with the following
properties:

Waves should represent large-scale global eigenoscillations of the
Sun. Their origin must be related to rotation, they must be strongly
anisotropic with respect to the azimuthal angle. Looking at the
characteristics of the solar cycle we immediately see the high coherency of
these global motions (the constant periods, phase shifts, amplitudes, the
latitude appearence, etc.). Activity grows in the first phase with a
timescale which is considerably shorter than the decay time in the second
phase; this fact and the quick eruptive release of energy by the reconnection
mechanism indicate that the waves must be unstable.

It is noteworthy that the inner gravity waves do not fulfill these
requirements. The quesion is whether the $r$-modes do?


\subsection{$r$-modes}.
    

In a non-rotating sphere ($\Omega=0$, where $\Omega$ is the angular frequency
of solar rotation) the wave motion is subdivided into two non-coupling
components: spheroidal $p-$, $f-$ and $g-$modes (for which the main restoring
forces are pressure gradient and buouancy) and toroidal modes (e.g. Unno
et al. 1989). Toroidal modes are degenerated horizontal eddy motions confined
to a spherical surface with a radius $r$ for which $\omega=0$,
$\dvrg{\vec{v}}=0$, and $\vec{v}=Q^m_l(r)\times
(0,\frac{1}{\sin\theta}\pd{}{\phi},-\pd{}{\theta}) Y^m_l(\theta,\phi)$. Here
$Y^m_l$ is the spherical harmonic with a degree $l$ and order $m$, $\theta$
is the colatitude, $\phi$ is the azimuthal angle in the spherical polar
coordinates, $\vec{v}$ is the fluid velocity field, $\omega$ is the angular
frequency of the fluid motion, and $Q^m_l(r) $ is an arbitrary amplitude
function. Toroidal modes have zero radial velocity but have non-zero radial
vorticity, $(\rot{\vec{v}})_r\ne 0$ (for the spheroidal modes it is vice
versa). These modes do not alter the equilibrium configuration.

When a slow rotation ($\Omega^2<\Omega^2_g=GM/R^3$) is included the spheroidal
modes are slightly modified but they keep their main properties. Degeneracy
of toroidal modes is removed only partially by the rotation, and
quasi-toroidal waves -- known as {\it r-modes} -- appear with a non-zero
frequency of $\omega\approx 2\Omega m/l(l+1)$ in the rotating frame
(Papaloizou \& Pringle 1978; Brayn 1889). Usually the governing equations of
the $r$-modes are obtained by expanding the initial physical variables of the
equations in the rotating system into power series with respect to the small
parameter $(\Omega/\Omega_g)^2$ ($\approx 10^{-4.7}$ for the Sun, e.g.
Papaloizou \& Pringle 1978; Provost et al. 1981; Smeyers et al. 1981; Saio
1982). These power series practically describe the deviation of the
surface of the star from its initial spherical state, resulting from rotation
through Coriolis and centrifugal forces. As a result of the deformation of
the spherical surface with a radius $r$ the radial vorticity of the toroidal
modes cause a surface pressure perturbation through the Coriolis force.
However, the $r$-modes practically keep the main properties of toroidal
flows: $v_r\approx 0$, $\dvrg{ \vec{v}} \approx 0$.
The degeneracy of the $r$-modes is that their frequencies hardly depend
on $Q^m_l(r)$, i.e. they are independent from the inner structure of the
star. For the $l=1$ modes the frequency in the inertial system is again close
to zero, $\omega\approx 0$ (Papaloizou \& Pringle 1978). The $r$-mode
equations define the amplitudes $Q^m_l(r)$, and taking into account the next
terms with small $\Omega^2/\Omega^2_g$ in the series practically does not
change the frequencies.

Due to the fact that the $r$-modes are practically surface deformation waves,
some similarity of these waves to the surface gravity waves or to the
$f$-modes is apparent.  For high $l$ the $f$-modes are an analogy to surface
gravity waves in a plane-parallel fluid with $\omega^2=gk_{\perp}$. In the
Cowling approximation $f$-modes with $l=1$ have zero frequency too,
$\omega\approx 0$ (Unno et al. 1989). This corresponds to a parallel
displacement of the whole star. For high $l$ the $f$-mode frequencies are
also independent of the inner structure, with $\omega^2\approx l\Omega^2_g$
(Gough 1980). So, $r$-modes are also fundamental rotating modes with an
inertial frequency, $\omega\le 2\Omega$.

For the Sun the properties of $r$-modes have been investigated in great
detail by Wolff et al. (1986) and Wolff (1998; 2000; and refs. therein).

Some properties of the $r$-modes are also similar to those of the Rossby
waves in geophysics (Pedlosky 1982). Similar to the Rossby waves and unlike
the $g$-modes the $r$-modes are strongly anisotropic. They propagate only in
azimuthal direction, opposite to rotation (i.e. they are retrograde waves in
the co-rotating frame). Because we are interested in length scales
corresponding to those of large sunspots, we have to
consider $r$-modes with $l\approx 100$. To get oscillations with periods of
years ($\omega/\Omega\approx 10^{-2}$) we must choose $m\approx l\gg 1$, just
such $r$-modes are physically more interesting (Lockitch \& Friedman 1999).
However, in the case of high $l$ the amplitudes of the $r$-modes will be
concentrated near the surface of the Sun (Provost et al. 1981; Wolff 2000),
and so they can actively interact with convective motions (Wolff 1997; 2000).
Because for these modes $v_r\approx 0$ and $\dvrg{\ \vec{v}}\approx 0$,
their chance to take part in the redistribution of angular rotation momentum
in the radiative interior is low. Note that the slow solar differential rotation does
not change the behavior of such $r$-modes with $m=l\gg 1$ (Wolff 1998).

Looking for further analogies between waves connected with gravity and with
rotation, we remember that beside the surface gravity waves there exist
internal gravity waves with $\omega^2\approx N^2k^2_x/(k_x^2+k_y^2)$, the
frequencies of which depend on the inner structure ($N$ is the 
Brunt-V\"ais\"al\"a frequency). Similar to these waves there exist `true'
Rossby (not deformation) waves, the frequency of which depends also on
the internal structure.


\subsection{Rossby waves}
   

We include here a short review on the main features of Rossby waves; they have
been investigated in great detail in geophysics (e.g. Pedlosky 1982; Gill 1982).
In the simplest case, that is in a plane-parallel, homogeneous, rotating
layer, the dispersion relation for the Rossby waves is $\omega\approx
2\Omega\beta k_x/(k_x^2+k_y^2+ k_z^2)$. Here $k_x$ is the wave number
perpendicular to the rotation axis, $k_z$ is expressed by the internal
deformation radius of Rossby which depends on the Brunt-V\"ais\"al\"a
frequency, $\beta$ is the transverse gradient (in $y$ direction) of the
Coriolis parameter: a vertical component of the `planetary' vorticity
$2\vec\Omega$ in the given local point. Unlike the $r$-modes the Rossby wave
frequencies are functions of the internal  structure and have maximum
dependence on the gradient $\beta$: $\omega\to 0$ if $k_x\to 0$ and if
$k_x\to\infty$. 

Any disturbance of the local flow in a rotating frame may generate waves of 
the Rossby type. These waves exist only if there is a gradient of the
potential vorticity $\vec\Pi =(\vec{\omega}_a\cdot\nabla\Xi)/\rho$.
Here an absolute vorticity is the sum of the relative and the planetary
vorticities, $\vec{\omega}_a=\rot{\vec{v}}+2\vec\Omega$, $\Xi$ is any
conserved scalar quantity, $\der{\Xi}{t}=0$ (for instance, for adiabatic
motion that could be the entropy or the density in the case of
incompressible plasma). The Rossby wave motion is a solution of the
nonlinear equation for transport of $\Pi$. The potential vorticity is
conserved if the medium is barotropic ($\nabla\rho\times\nabla p=0$) and if
there are no torques. The rotation of the frame is added to any vorticity in
the velocity field. Any motion within a rotating fluid serves as a potential
source for vorticity. 

The relative vorticity may be evoked by the geometrical surface as well as by
internal gradients. It depends on the choice of the function 
$\Xi(r,\theta,\phi)$ and on $\vec{\Omega}(r,\theta)$. For example, an
unevenness of the ocean bottom causes the topographic Rossby waves, or a
dependence of the Coriolis parameter on the earth latitude
($F=2\Omega\sin\varphi$, where $\varphi$ is the geographic latitude)
is the main cause of atmospheric Rossby waves.

In the solar dynamo context the ability of Rossby waves to induce
solar-like magnetic fields has been considered by Gilman (1969). Here the
mechanism
for sustaining the Rossby waves is a latitudinal temperature gradient in a
thin, rotating, incompressible convective zone.  To interpret the dynamical
features of large-scale magnetic fields the Rossby vorticies excited within a
thin layer beneath the convective zone are considered by Tikhomolov \& Mordvinov (1996)
as the result of
a deformation of the lower boundary of the convective zone.


\subsection{$R$-modes}
   

From the discussion in Subsection 1.4 we conclude that just Rossby-like waves
could be suitable for our requirements. As the main driving mechanism  we
choose a latitudinal (or horizontal) differential rotation,
$\Omega=\Omega(\theta)$. Baker \& Kippenhahn (1959) have pointed out
that the uniform rotation of a star is not a typical case. Low frequencies
(periods of years) could easily be obtained searching for the
eigenoscillations of the Sun's radiative interior, where the gradient of the
rotation speed is close to zero (in accordance with the helioseismology
results). Large scales such as those associated with sunspots
($k_xR_{\odot}\sim 100$) decrease the frequencies too. Similar to the
$r$-modes the Rossby waves are strongly anisotropic (retrograde waves), but
unlike the $r$-modes these waves are concentrated close to the solar center.
These results have already been obtained by Oraevsky \& Dzhalilov (1997), who
investigated the trapping of adiabatic, incompressible Rossby-like waves in
the solar interior. In the present work we take into account compressibility
for the nonadiabatic waves. We look for unstable waves. It is clear that the
necessary condition for the Kelvin-Helmholtz shear  instability,
$4N^{2}-(r\der{\Omega}{r})^{2}<0$ (Ando 1985), is not fulfilled in the
radiative interior. Then we decided to include the thermal $\en$-mechanism of
instability which is favoured at low frequencies (Unno et al. 1989). To
balance the $\en$-mechanism the radiative losses in the diffusion regime are
included. To exclude all geometrical effects we ignore the influence of the
spherical surface at the given radius. The dispersion relation in the limit
of adiabatic incompressiblity and at very low frequencies is the same as that
for Rossby waves in geophysics. In order to distinguish these rotational body
waves from the $r$-modes we call them {\it R-modes} (Rossby rotation).

The governing fourth order equation is obtained from the basic equations in  
Sect.\,2. Some qualitative analysis of the wave cavity trapping is done 
for the simpler adiabatic case in Sect.\,3. Using the asymptotic solutions
obtained in Sect.\,4 the complex boundary value problem is solved in Sect.\,5.
The calculation of the eigenfrequencies and the instability analysis are
done in Sect.\,6. The obtained unstable modes are shortly discussed in 
Sect.\,7.  


\section{Setting the problem}
   
   \subsection{Basic equations}
   

Let us investigate global motions with large timescales such that the Rossby
number is small, $\omega/\Omega < 1$. Before the appearence of any
disturbances the basic stationary state of the rotating star is defined
mainly by the balance of pressure gradient, gravity force, and forces
exerted by the noninertiality of the motion (Coriolis and centrifugal forces).
In the case of an incompressible fluid with a homogeneous rotation rate
usually this state is called `geostrophic balance'. A star disturbed by an
external force tends to return to this basic state. Our aim is to study for
the Sun the dynamics of small deviations from the steady geostrophic balance.
For this purpose it is natural to write the dynamic equations in a frame
rotating together with the Sun. The magnetic field will be ignored. For
arbitrary $\Omega(r,\vartheta)$ the equation of momentum in conventional
definitions is given by   \be
\label{mot1}
\der{\vec{v}}{t} + 2\vec{\Omega}\times\vec{v} = 
-\frac{1}{\rho}\nabla{p} + \vec{g} + \vec{r}\times\der{\vec{\Omega}}{t} - 
\vec{\Omega}\times(\vec{\Omega}\times\vec{r})+ \mu_v\nabla^2\vec{v} ,
\ee
where $\mu_v$ is the kinematic/turbulent viscosity coefficient. In the next
steps this equation will be simplified keeping the main features of the motion.
We consider linear waves without taking into account convective and
meridional flows, $\vec{v_0} = 0 $, and we suppose that the angular velocity 
$\Omega $ does not depend on time. For the basic state we have from 
Eq.~(\ref{mot1}) 

\[ 
-\frac{1}{\rho_0}\nabla{p_0} + \vec{g} = 
\vec{\Omega}\times(\vec{\Omega}\times\vec{r}). 
\]
However, we can exclude practically everywhere in the Sun
centrifugal acceleration and consider it as a small correction to $\vec{g}$.
Then the  spherically symmetric basic state is defined by
$\nabla{p_0}\approx\rho_0\vec{g}$.
For the linear oscillations we have from Eq.~(\ref{mot1}):
\begin{eqnarray}
\label{mot2}
\pd{\vec{v}}{t} + 2\vec{\Omega}\times\vec{v} + 
r\sin\vartheta(\vec{v}\cdot\nabla\Omega)\vec{e_{\varphi}} & = & \nonumber \\
-\frac{1}{\rho_0}\nabla{p^\prime} + \vec{g}^{\prime} + 
\frac{\nabla p_0}{\rho_0^2}\rho^\prime + \mu_v\nabla^2\vec{v} . & &
\end{eqnarray} 
Here $\vec{e_{\varphi}}$ is the unit vector in azimuthal direction, and
variables with a prime are Eulerian perturbations. Eq.~(\ref{mot2}) is
written in rotating spherical polar coordinates ($r,\vartheta,\varphi$).    
It coincides with the equation of motion by Unno et al. (1989) which is
derived for an inertial frame, if the operator $\pdtext{}{t}$ is replaced by
$\pdtext{}{t} +  \Omega\pdtext{}{\varphi}$. To simplify our further
discussion we shall use the Cowling approximation, $\vec{g}^{\prime}=0$,
which has a sufficient degree of accuracy for the analysis of short waves.

The next simplification of Eq.~(\ref{mot2}) is connected with the quasi-rigid
rotation of the inner part of the Sun below the convective zone, which
is known from the solution of the inverse problem of helioseismology. In this
case we can omit the third term of the l.h.s of Eq.~(\ref{mot2}). Such a 
restriction of the gradients of $\Omega(r,\vartheta)$ requires to obey the
conditions:
\be
\label{grad}
2\Omega\gg r\pd{\Omega}{r},\;\; 2\Omega\gg \tan(\vartheta)\pd{\Omega}{\vartheta} .
\ee
To obtain these conditions we used the components of the vector $\vec\Omega$
in spherical polar coordinates, $\vec\Omega = \{\Omega\cos\vartheta,$
$-\Omega\sin\vartheta, 0\}$. 

The next equations are the mass and energy conservation equations in the
standard form:
\begin{eqnarray}
\label{mass}
\der{\rho}{t}+\rho{\,}\func{div}{\vec{v}}&=&0 ,\\
c_v\rho\der{T}{t} + p{\,}\func{div}{\vec{v}} & = & - {\cal{L}} ,\label{ener}\\
\mbox{where:} - {\cal{L}} = Q(\rho,T) &-& \nabla\cdot\vec{q}_{R}-
 \nabla\cdot(K_0T^{5/2}\nabla T) . \nonumber
\end{eqnarray}
Another form of Eq.~(\ref{ener}) is 
\be
\label{ener0}
\der{p}{t} - c_s^2\der{\rho}{t}=-(\gamma-1){\cal{L}} .
\ee
Here $\gamma=c_p/c_v$, $c_p$ and $c_v$ are the specific heats at constant
pressure and volume, respectively, $c_s$ is the sound speed, the source
function $Q(\rho,T)$ is the sum of nuclear and viscous heat generation rates
per unit volume: $\rho(\en_N + \en_v)$, $\vec{q}_{R}$ is the radiative energy
flux, $K_0T^{5/2}$ is the coefficient of electron heat conductivity. In the
interior of the Sun we shall neglect viscous heating ($\en_v =0$). For
the power of nuclear reactions we have  $Q\approx\rho^2 T^{\alpha}$. In
particular for the p-p reaction $Q=\rho\en_{pp} \approx 
9\times 10^{-30}\chi^2_H\rho^2 T^4$, where $\en_{pp}$ is given in
[erg/g\,s] and $\chi_H=0.73$.    

In the limit of incompressible fluid, $\dertext{\rho}{t}=0$ or
$\func{div}{\vec{v}}=0$, it follows from Eq.~(\ref{ener0}) that the condition
$c_s^2\to\infty$ is not needed to satisfy $\dertext{p}{t}\neq 0$. Hence, in a
dissipative (${\cal{L}}\ne 0$),  incompressible fluid sound cannot propagate
instantaneously. It means,we  cannot use the condition of $c_s^2\to\infty$ to
get the incompressible limit for nonadiabatic waves .
 
Now we will try to simplify the energy loss function ${\cal{L}}$ assuming 
reasonable approximations for the Sun's interior. We use the formula for the
heat conductivity of a fully ionized gas to show that in the Sun radiative
transport of energy is more important than that by particle heat
conductivity:
\[
\kappa_T = K_0 T^{5/2}\approx 10^{-6}T^{5/2}\,\,{\rm erg/s\,cm\,K} .
\]
The radiative flux is given by the radiative diffusion equation
\be
\vec{q}_R = -\kappa_R\nabla T, \ \ \kappa_R =\frac{16\sigma_S}{3\chi}T^3,
\ee
where $\kappa_R$ is the radiative heat conductivity, $\sigma_S = a_R c/4 = 
5.67\times 10^{-5}$\,erg/cm$^2$\,K$^4$\,s is the Stefan-Boltzmann
constant, and $\chi$ is the Rosseland mean opacity. For the opacity we will
use Kramers' formula: $\chi=3.68\times10^{22}\rho^2 T^{-7/2}$\,cm$^{-1}$.
Then, for the radiative conductivity we have
\[
\kappa_R \approx 8.22\times 10^{-27}\frac{T^{13/2}}{\rho^2} 
\,\,{\rm erg/s\,cm\,K} .
\]  
Now we find that $\kappa_R/\kappa_T\approx 10^{-20}T^4/\rho^2 > 10^3$.
This condition is fulfilled in the whole Sun. That means, we can exclude the
heat conductivity term from Eq.~(\ref{ener}). The next simplification is
connected with the ratio $T^{5/2}/\rho$ which is practically not changed over
the radius and $\approx 10^{16}$. It means, we can introduce the constant 
$\kap_0=8.22\times 10^{-27}T^5/\rho^2\approx 8.2\times 10^5$. Then we have
\[
\kappa_R\approx \kap_0 T^{3/2}\,\,{\rm erg/s\,cm\,K} .
\] 
Now for the right-hand side of Eq.~(\ref{ener}) we have
\be
\label{L}
-{\cal{L}} \simeq\frac{2}{5}\kap_0\nabla^2 T^{5/2} + Q.
\ee
The main non-perturbed energetic state of the Sun is defined by the condition
${\cal{L}}_0=0$. In our case this condition is
\be
Q_0 = -\frac{2}{5}\kap_0\frac{d^2 T_0^{5/2}}{dr^2}.
\ee


   \subsection{Choice of the frame of reference}
      

In the next step we try to get the analytical solutions of the wave
equations and to solve the boundary value problem. For this aim we shall
investigate the short waves (WKB) approximation for which the effects of
curvature are unimportant. It means we can apply the rotating 
plane-parallel stratification approximation. Such an approach has its
advantages and disadvantages. The main advantage is that at the end points of
the integration path (center and pole of the Sun) we have no singularity, and
this gives a chance to find the solutions analytically. The disadvantages are
connected with the following: a) long waves are excluded but they are very
important, for instance, for the transfer of angular momentum; b) the lost
`end' singularities correspond to a real physical behavior of the waves;
c) in this approach we get two distinct directions, $\Omega$ and $g$
(taking into account all components of $\vec\Omega$), which are absent in a
self-gravitating, radially stratified sphere. The deviation of the direction
of stratification of the plane-parallel fluid layer from the direction of its
axis of rotation should lead to additional, physically doubtful results. In
geophysical hydrodynamics this problem is solved by applying the
`$\beta$-plane' within the frame of the traditional approximation (Pedlosky
1982; Shore 1992), where the component of $\vec{\Omega}$ parallel to
$\vec{g}$ in the given surface point (local vertical) is retained only in the
governing equations. Here we will use the same approach.

Let us take an arbitrary point at the surface of the rotating sphere. The
position of this point is defined by its radius $r$, its co-latitude
$\vartheta$, and its azimuth angle. We assign to this point a local,
left-handed Cartesian system of coordinates $\{x,y,z\}$, where the $\vec
z$-axis is directed along the radius (local vertical), the direction of the
$\vec y$-axis is meridional (towards the pole), and that of the $\vec x$-axis
azimuthal. In this frame of reference $\vec\Omega = \{0,\Omega_y,\Omega_z \}
= \{0, \Omega\sin\vartheta, \Omega\cos\vartheta\}$. Strictly speaking, the
$\vec z$-axis coincides with the rotation axis only at the pole
($\vartheta=0$). In the case of a homogeneous fluid $\Omega$ is included into
the wave equation only in the term \[
(\vec\Omega\nabla)^2 = \Omega_y^2\frac{\partial^2}{\partial y^2}+
2\Omega_y\Omega_z\frac{\partial^2}{\partial y\partial z}+
\Omega_z^2\frac{\partial^2}{\partial z^2} .
\]
Here we can neglect $\Omega_y$ ($\Omega_y=0$), if the condition 
$|\Omega_y\pd{}{y}|\ll|\Omega_z\pd{}{z}|$ is fulfilled. This condition
is named `traditional approximation'. For harmonic motions we have
$\pd{}{y}=ik_y$, $\pd{}{z}=ik_z$, and the traditional approximation
corresponds to $|k_y|\ll|k_z|\cot\vartheta$. If the spatial scale
of the wave motion in vertical direction is much smaller than in horizontal
direction (at latitudes not close to the equator), we can restrict ourselves
to retain only $\Omega_z$ in the governing equations. This condition for the
traditional approximation remains valid, if a radial stratification is
included (Lee \& Saio 1997).

In order to construct the `$\beta$-plane' limit we expand $\Omega_z$ around
a fixed $\vartheta_0$ ($\vartheta=\vartheta_0 + \delta\vartheta$):
$ \Omega_z=\Omega(\vartheta)\cos\vartheta\approx
(1+\beta y)\Omega_0\cos\vartheta_0 $. Here $y=R_{\odot}\delta\vartheta$,
$\Omega_0=\Omega(\vartheta_0)$, and
\[
\beta=\left(\frac{1}{\Omega_0}\pd{\Omega_0}{\vartheta}-\tan\vartheta_0\right)
\frac{1}{R_{\odot}} .
\]
Rossby waves are possible only if $\beta\ne 0$. The parameter $\beta$
is a sum of two terms: the second one ($\tan\vartheta$) is due to the
geometrical change of the Coriolis parameter with latitude. This
term exists always, even if the rotation is rigid. The $r$-modes are
connected with this term. The first term in $\beta$ appears if the
differential rotation is considered, and the $R$-modes are connected with
this term. Close to the pole ($\vartheta_0\to 0$) the first term is dominant.

Note that such a `$\beta$'-limit is applicable also around the equator plane,
where the traditional approximation does not fit. The advantage of this limit
is the possibility to include $\vartheta$ as a parameter in the Cartesian
system. In this way we use here an inertial Cartesian system of coordinates
$(x,y,z)$ in a frame rotating with an angular frequency $\Omega(y,z)$. All
non-perturbed model variables are functions of $z$ only, and for the gravity
acceleration we have $\vec g=\{0,0,-g(z)\}$.
For the observer from the non-rotating frame 
the elements of fluid are moving due to rotation with a velocity $\vec{V_0} =
\vec{\Omega}\times\vec{r}=\{-\Omega y,\Omega x,0\}$, where in the frame of
our approximation $\Omega \approx \Omega_z$. $V_{0x}<0$ means that the
$x$-axis is directed opposite to the rotation.


   \subsection{Oscillation equations}
      

For linearization each physical variable $f=f_0+f'$ is decomposed into a mean
term $f_0$ and a small fluctuating term $f'$. Neglecting terms of higher
order than the first one we get our oscillation equations 
\be \pd{\rho'}{t} + v_z\der{\rho_0}{z} + \rho_0{\,}\func{div}{\vec{v}}=0,
\label{mas2} 
\ee
\be
\pd{\vec{v}}{t}+2\vec{\Omega}\times\vec{v} =  
-\frac{1}{\rho_0}\nabla{p'} + \vec{g}\frac{\rho'}{\rho_0} + 
\mu_v\nabla^2\vec{v}, \label{mot3} 
\ee
\[
c_v\rho_0\left(\pd{T'}{t}+v_z\der{T_0}{z}\right) +
p_0{\,}\func{div}{\vec{v}}= \kap_0\nabla^2\left(T_0^{5/2}\frac{T'}{T_0}\right) + 
\]
\be
+ Q_0\left(2\frac{\rho'}{\rho_0} +
\alpha\frac{T'}{T_0}\right) = -{\cal{L}}'. \label{ener2}
\ee
For adiabatic oscillations ${\cal{L}}'=0$. We approach this regime by
setting formally $\kap_0=0$ (Eq. 9). But in the nonadiabatic case, which is
considered here, ${\cal{L}}'$ is the sum of radiative damping and the
$\en$-mechanism terms. Eqs.~(\ref{mas2})-(\ref{ener2}) must be completed then
by the equation of state, $p=p(\rho,T)$. For an ideal gas we have
\be
\label{idl}
\frac{p'}{p_0} = \frac{\rho'}{\rho_0} + \frac{T'}{T_0} .
\ee
For the ideal gas $p_0=R_g\rho_0 T_0/\mu_m$, where $R_g$ is the gas constant,
$\mu_m$ is the molecular weight, and the squared adiabatic sound speed $c_s^2=
\gamma c^2$ is defined by the squared isothermal sound speed $c^2=
p_0/\rho_0$. 

We define
\begin{eqnarray}
\tilde{Q_0}&=&\frac{Q_0}{c_v\rho_0 T_0}, \ \ \ \tilde{\kap_0}=\frac{\kap_0}
{c_v\rho_0 T_0}, \ \ N^2 = g\left(\frac{1}{\gamma}\kap_p -
\kap_{\rho}\right),\nonumber\\
\kap_T&=&\frac{1}{T_0}\der{T_0}{z}, \ \ \
\kap_{\rho}=\frac{1}{\rho_0}\der{\rho_0}{z},\nonumber\\
\kap_p&=&\frac{1}{p_0}\der{p_0}{z} = \kap_{\rho} +\kap_T = -\frac{g}{c^2} .
\nonumber
\end{eqnarray}
Here $N^2(z)$ is the squared Brunt-V\"ais\"al\"a frequency. $\kap_p,
\kap_{\rho}$, and $\kap_T$ are the reciprocal pressure, density, and
temperature inhomogeneity scales, respectively, and $\tilde{Q_0}$ is the
reciprocal of the characteristic Kelvin-Helmholtz time-scale --- a deviation
from the thermal balance of the star is restored
during this time. As the coefficients of Eqs.~(\ref{mas2})-(\ref{idl}) are
independent of the time $t$ and of the space coordinate $x$ we can set 
\be
\pd{}{t} = -i\omega, \,\,\, \pd{}{x} = ik_x . 
\ee 
Now we exclude $\rho'$ from
the system of Eqs.~(\ref{mas2})-(\ref{idl}) and have
\begin{eqnarray}
-i\omega\left(\frac{p'}{p_0}-\frac{T'}{T_0}\right) +\kap_{\rho}{\,}v_z +
u&=&0,\label{9} \\
-ik_xc^2\frac{p'}{p_0} +2\Omega{\,}v_y + \check D v_x&=&0, \label{10}  \\
-c^2\pd{}{y}\frac{p'}{p_0}-2\Omega{\,}v_x + \check D v_y&=&0, \label{11}\\
-c^2\pd{}{z}\frac{p'}{p_0} +g\frac{T'}{T_0}+\check D v_z&=&0, \label{12}\\
-(i\omega+2\tilde{Q_0})\frac{p'}{p_0}+\kap_p v_z +\gamma u&=&\check f,\label{13}\\
u=\func{div}{\vec v} = ik_x v_x+\pd{v_y}{y}+\pd{v_z}{z},\label{8}
\end{eqnarray}
where the operators $\check D$ and $\check f$ are defined as
\be
\check D=i\omega+\mu_v\nabla^2,\,\, \check f=(\alpha-2)\tilde{Q_0}\frac{T'}{T_0}+
\tilde{\kap_0}\nabla^2\!\left(T_0^{5/2}\frac{T'}{T_0}\right), \label{f}
\ee
and the relation $p_0/(c_v\rho_0T_0)=\gamma-1$ is used.
In the next steps the viscosity appears in the coefficients of the equations
in the form
$\frac{c^2}{\omega}+\mu_v\cdot O(1)$. For a fully ionized plasma the kinematic
viscosity is $\mu_v\approx 10^{-16}T^{5/2}/\rho$ [cm$^2$/s]. For the solar
situation $\mu_v\approx O(1)$, if we do not take into account the turbulent
viscosity. As we are interested in very low frequencies the condition 
$\frac{c^2}{\omega}\gg\mu_v$ is valid. Then we can put $\check D\simeq
i\omega$ (non-viscous case). Now Eqs.~(\ref{10})-(\ref{11}) define the
horizontal components of velocity and hence its
two-dimensional divergence  $\func{div}_{\perp}\vec{v}$:
\begin{eqnarray}
v_x&=&-\frac{k_xc^2}{\omega}\frac{\sigma^2}{1-\sigma^2}
\left(1+\frac{1}{\sigma k_x}\pd{}{y}\right)\frac{p'}{p_0} ,\\
v_y&=&i\frac{k_xc^2}{\omega}\frac{\sigma}{1-\sigma^2}
\left(1+\frac{\sigma}{k_x}\pd{}{y}\right)\frac{p'}{p_0}, \\
\func{div}_{\perp}\vec{v}&=&
c^2ik_x\frac{\sigma'}{\omega}\check{M}\frac{p'}{p_0},
\end{eqnarray}
where
\[
\check{M}=1+\frac{2\sigma}{k_x}\pd{}{y}+\frac{\sigma^2}{k_x\sigma'}
\left(\frac{\partial^2}{\partial y^2}-k_x^2\right),\,\,
\sigma=\frac{\omega}{2\Omega},\,\,\sigma'=\pd{\sigma}{y}.
\]
Excluding $u=\func{div}_{\perp}\vec{v}+\pdtext{v_z}{z}$ by using  
Eq.~(\ref{9}) we have the system of equations
\begin{eqnarray}
i\omega{\,}v_z = c^2\pd{}{z}\frac{p'}{p_0}-g\frac{T'}{T_0}, \label{vz}\\
\frac{T'}{T_0}-\check{s}(y,z)\frac{p'}{p_0}+\frac{1}{i\omega}\left(\kap_{\rho}+
\pd{}{z}\right)v_z=0,\\
q\frac{p'}{p_0} + \frac{1}{i\omega}\frac{N^2}{g}v_z = 
\frac{T'}{T_0}+\frac{1}{\gamma_*}\check{f} \label{p'},
\end{eqnarray}
where the dimensionless quantites are
\be
\check{s}=1-c^2k_x\frac{\sigma'}{\omega^2}\check{M},\,
\gamma_*=\gamma i\omega,\,\, 
q=\frac{\gamma-1}{\gamma}-\frac{2\tilde Q_0}{\gamma_*} .
\ee
Now we use the condition $\sigma\ll 1$ and receive
\begin{eqnarray}
v_x &=& -\frac{c^2}{2\Omega}\pd{}{y}\frac{p'}{p_0}, \label{vx} \\
v_y &=& \frac{c^2}{2\Omega}ik_x\frac{p'}{p_0} . \label{vy}
\end{eqnarray}
In the case of rigid rotation, $\pd{\Omega}{y}=0$, it follows immediately from
Eqs.~(\ref{vx})-(\ref{vy}) that $\func{div}_{\perp}\vec{v}=0$. It means,
that the incompressible case ($\func{div}\vec{v}=0$) for which everywhere
$v_z=0$ (from the boundary condition at the center) is not so interesting for
astrophysical situations. In this case $\rho'=0$ and the equation of heat
conductivity, Eq.~((\ref{ener2}), is separated from the equation of motion. In
geophysical situations just this case is interesting, when geostrophic eddies
are investigated. For our task waves with $v_z\ne 0$ are more important.

Our next step is to separate the $z$ and $y$ dependence of the variables in
the  governing equations to have finally one ordinary differential equation.
In the general case such a separation is not possible, and we consider here
the very simple case when the function $\pd{}{y}(\frac{1}{\Omega})$ is 
independent of $y$. Only in this case it is possible to separate the
equations with respect to the variables $y$ and $z$ and we can write
$\frac{\partial^2}{\partial y^2}=-k_y^2$. Here $k_y$ should be determined
from the boundary conditions, and a complex $k_y$ is not excluded. In this
way the system of partial differential equations in the plane-parallel
approximation is reduced to ordinary differential equations. Now we assume
the following formula for the rotation profile
\be
\Omega(y,z)=\frac{\tilde\Omega(z)}{1+\beta y/R_{\odot}},\label{OM}
\ee 
where $\beta>0$. This profile reveals that the rotation rate decreases at the
pole. If $y\to\infty$ then $\Omega\to 0.$ Helioseismology inversions predict
$\Omega(\vartheta)$ in the convective zone which is almost the
same as that at the surface ($\beta\approx O(1)$), but below the bottom of
the convective zone (at the tachocline) rotation is close to a solid-body
rotation ($\beta\ll 1$). As $y_* = y/R_{\odot} \le 1$, in the deepest layers
of the Sun we can assume the rotation rate $\Omega(y,z)\approx\tilde\Omega(z)
(1-\beta y_*)$, where $\tilde\Omega(z)\approx O(\Omega_{\odot})$.   
Then we have in Eq. (28)
\be
\check{M}\approx M= 1+\frac{\omega k^2_{\perp}}{k_x2\Omega^\prime(y)},
\,\,\check{s}=s\approx 1-M\frac{c^2k_x}{2\omega}\pd{}{y}\frac{1}{\Omega} ,
\ee
where $k^2_{\perp} = k^2_{x} + k^2_{y}$ .

Our next step is to derive one differential equation for the temperature
perturbations. The variable $v_z$ is easily excluded from
Eqs.~(\ref{vz})-(\ref{p'}). Then we get for the pressure perturbations
\be
\label{ap'}
a\frac{p'}{p_0}=\frac{T'}{T_0}+\left(\frac{\omega}{cd}\right)^2q\left[\left(1
-\frac{N^2}{\omega^2}\right) \frac{T'}{T_0}+\frac{\check f}{\gamma_*}\right] + 
\ee
$$
+ \left(\kap_{\rho}+\pd{}{z}\right)
\left(\frac{T'}{T_0}+ \frac{\check f}{\gamma_*}\right)\frac{1}{d(z)},
$$
where
\be
a=s+\left(\frac{\omega q}{cd}\right)^2 + \left(\kap_{\rho}+
\pd{}{z}\right)\frac{q}{d}, \ \
d=\frac{N^2}{g}=\frac{\kap_p}{\gamma}-\kap_{\rho} .
\ee
The parameter $d$ is positive in the solar radiative interior, $d > 0$,
while we have $d\le 0$ in the convective zone. The pressure perturbations
(Eq. 33) have a singularity at $a=0$. However, it will be shown below that
this singularity is removable.

Introducing a new dependent variable
\be
\Theta =
\kappa\frac{T'}{T_0},\,\,\,\kappa=\left(\frac{T_0}{T_{00}}\right)^{5/2},
\,\,T_{00}={\rm const} ,  \ee
we receive the final equation of fourth order
\be
\label{eq4}
\Theta'''' + A_3\,\Theta''' + A_2\,\Theta''+ A_1\,\Theta'+A_0\,\Theta=0,
\ee
where $\Theta'=\dertext{\Theta}{z}$,$\,\en_0=\tilde\kap_0 T_{0}^{5/2}/\gamma_*$ and
\begin{eqnarray*}
A_0&=&a_1a_2+a_2'-\frac{a\omega^2}{\en_0c^2}\left(\frac{N^2}
{\omega^2}-\en_0a_3\right), \hspace{18mm} (36')\\
A_1&=&a_1a_3+a_2+a_3', \\
A_2&=&a_3 - a_1 b_2 - b'_2, \ \  A_3 =-\der{}{z}\func{ln}(p_0T_0ad^2),\\
-a_1&=&\frac{a'}{a}+\frac{d'}{d}+\kap_p +\frac{q\omega^2}{dc^2}, \\
a_2&=&\frac{1}{\en_0}\left(b_1-\frac{5}{2}\kap_T\right)+\kap_1'+b_2\tilde{k^2_{\perp}},\\
a_3&=&\frac{1}{\en_0}-\tilde{k^2_{\perp}}, \ \ 
\tilde{k^2_{\perp}}=k_x^2 + k_y^2 -\kap_1=k^2_{\perp}-\kap_1,\\
b_1&=&\frac{\kap_p}{\gamma}-\frac{d'}{d}-\frac{q\omega^2}{dc^2}
\left(\frac{N^2}{\omega^2}-1\right), \ 
b_2=\kap_T+\frac{d'}{d}-\frac{q\omega^2}{dc^2}, \\
\kap_1&=&\frac{2}{5}(2-\alpha)\frac{\kappa''}{\kappa}, \ \  
q=\frac{\gamma-1}{\gamma}+\frac{4}{5}\en_0\frac{\kappa''}{\kappa} .
\end{eqnarray*}

In order to solve Eq.~(\ref{eq4}) we normalize it to get a dimensionless
equation. For this purpose we normalize the radial distance to the solar
radius:
$z\Rightarrow z/R_{\odot}$ and get finally the following equation
\[
\en\Theta''''+\en\varphi_3\Theta'''+(1+\en\varphi_2)\Theta''+
\]
\be
\label{weq}
+(\psi_1+\en\varphi_1)\Theta'+(\psi_0+\en\varphi_0)\Theta=0,
\ee
where we kept the notation $\tilde{k_{\perp}^2}\Rightarrow 
R_{\odot}^2\tilde{k^2_{\perp}}$, the inverse scale heights $\kap_{p,T}$ are
defined as above (below Eq.~(13)) but with a normalized $z$, and
\begin{eqnarray*}
\psi_0&=&-a\tilde{\omega}^2(N^2/\omega^2-1+\en \tilde{k_{\perp}^2})+
b_1(a_1+\kap_T + b_1'/b_1),  \\
\psi_1&=&-\der{}{z}\func{ln}(T_0^{3/2}ad^2), \ \ 
\varphi_3=-\der{}{z}\func{ln}\left(p_0T_0ad^2\right),\\
\varphi_0&=&a_1(\kap_1'+b_2\tilde{k_{\perp}^2})+(\kap_1'+
b_2\tilde{k_{\perp}^2})', \hspace{22mm}  (37') \\
\varphi_1&=&-\tilde{k_{\perp}^2}\varphi_3 + 
\frac{4}{5}\kap_p\frac{\kappa''}{\kappa}+2\kap_1', \\
-\varphi_2&=&\tilde{k_{\perp}^2}+a_1b_2+b_2'-\tilde{\omega}^2a, \\
-a_1&=&\frac{a'}{a}+\frac{d'}{d}+\kap_p+\tilde{\omega}^2\frac{q}{d}, \ \ 
q=\frac{\gamma-1}{\gamma}+\frac{4}{5}\en\frac{\kappa''}{\kappa},  \\
b_1&=&\frac{\kap_p}{\gamma}-\frac{5}{2}\kap_T-\frac{d'}{d}-
\tilde{\omega}^2\frac{q}{d}\left(\frac{N^2}{\omega^2}-1\right), \\ 
b_2&=&\kap_T +\frac{d'}{d}-\tilde{\omega}^2\frac{q}{d}, \ \ 
M=1-\frac{\omega}{2\tilde\Omega}\frac{\bar{k^2_{\perp}}}{\beta\bar{k_x}} ,\\
a&=&1-\Lambda+\left(\tilde{\omega}\frac{q}{d}\right)^2+\left(\kap_{\rho}+
\pd{}{z}\right)\frac{q}{d},\\
\en&=&\frac{\en_0}{R_{\odot}^2}=\frac{\kap_0T_0^{5/2}}{c_v\rho_0T_0}
\frac{1}{\gamma_* R_{\odot}^2}=\frac{1}{\gamma_* R_{\odot}^2}
\frac{\kappa_R}{c_v\rho_0}\\
\bar{k}_{x,\perp}&=&k_{x,\perp}R_{\odot}, \ \ \tilde{\omega}=\omega
R_{\odot}/c,\ \ \Lambda=\beta\frac{k_xc^2M}{2\tilde{\Omega}\omega R_{\odot}} .
\end{eqnarray*}
Remind that $\gamma_* =\gamma i\omega$. Here the main parameter is $\Lambda$,
it includes the rotation rate gradient ($\beta$=const).

Because we are interested in very low frequency oscillations with periods of
1--20 years, we take $\omega\approx 10^{-8}\,s^{-1}$, $\frac{T_0^{5/2}}
{\rho_0}\approx {\rm const}=10^{16},\,c_v\approx 2\times 10^8,\,\kap_0\approx
8\times10^5$, and we have a small parameter for our task
\be
\en(z)\approx 10^{-8}\frac{T_{\rm c}}{T_0(z)}.
\ee
Here $T_{\rm c}$ is the temperature of the solar center. For the whole Sun
this parameter $\en$ is changed in the interval $10^{-8}\le\en\le 10^{-5}$.
$\en$ characterizes the degree of nonadiabaticity of the waves, it is defined
as the ratio of the wave period to the reciprocal of the Kelvin-Helmholtz
time.


\section{Adiabatic case}


For idealized adiabatic waves ($\en=0$) we have a second order equation,
\be
\Theta''+\psi_1\Theta'+\psi_0\Theta=0 .
\ee
Introducting a new variable it may be written in standard form
\be
\label{Y}
\Theta=Y\,\sqrt{T_0^{3/2}ad^2},\ \ \ Y''(z)+I(z)Y=0 ,
\ee
where
\be
\label{I}
I=\psi_0-\frac{1}{4}\left[\der{}{z}\func{ln}(T_0^{3/2}ad^2)\right]^2 +
\frac{1}{2}\frac{d^2}{dz^2}\func{ln}(T_0^{3/2}ad^2) .
\ee
$\psi_0$ and $a$ are defined by Eqs. (37') and $d=R_{\odot}N^2/g$.
The behavior of the function $I(z)$ gives a possibility to analyze
qualitatively the waves in the solar interior and the boundaries from
which the waves are reflected and become trapped. If $I>0$ we have oscillating
solutions and if $I\le 0$ the waves are exponentially decreasing
(evanescent) with $z$. Now we shall breafly discuss the incompressible and
compressible cases.

As we consider adiabatic waves, the transition to the {\it{incompressible}} 
case can be done by $c^2\to\infty$. In this case $\kap_p\to 0,\,
d\to -\kap_{\rho},\,|\Lambda|\to\infty$ and instead of Eq.~(\ref{I}) we
obtain \be
\label{Inc}
I\approx \beta\bar{k}_x\frac{\omega}{2\tilde{\Omega}}\left(\frac{N^2}{\omega^2}-
1\right)-\frac{5}{2}\kap_T\kap_{\Omega}-\frac{1}{4}\left(\frac{5}{2}\kap_T-
\kap_{\Omega}\right)^2 ,
\ee
where $\bar{k}_x=k_xR_{\odot}$ and $\kap_{\Omega}=\dertext{\func{ln}
\tilde{\Omega}}{z}$. Here there are three possibilities:\\
1) absolute rigid body rotation, $\beta=\kap_{\Omega}=0$. In this case $I<0$
and we have no cavity solutions;\\
2) rigid rotation with respect to latitude ($\beta=0$), but
`vertical' differential rotation with respect to the radius
($\kap_{\Omega}\ne 0$).  It follows from Eq.~(\ref{Inc}) that in this case
for oscillating solutions we must have $\kap_{\Omega}>0$ as $\kap_T<0$ is
obeyed in the inner part of the Sun. It means that solar rotation must have a
decreasing speed toward the center ($\dertext{\Omega}{z}>0$). Then for $I>0$
the condition  $10|\kap_T|\kap_{\Omega}>(5\kap_T/2-\kap_{\Omega})^2$ must be
fulfilled;\\
3)The most realistic case for the Sun is a differential rotation in both 
directions ($\beta\ne 0$ and $\dertext{\Omega}{z}\ne 0$). In this case we
have various chances to get trapped waves. We are interested in
oscillations with $\omega\approx 10^{-8}\times s^{-1}$ (years) and 
$\bar{k}_x\approx 10^2$ ($\lambda\approx 3\times 10^4$\,km, 
e.g. sunspots). For solar conditions ($\Omega_{\odot}\approx 2.86\times
10^{-6}$\,s$^{-1}$ and $N^2_{\rm max}\approx 6\times 10^{-6}$\,s$^{-2}$) we
have $\bar{k}_x\omega/2\Omega\approx 1$ and $N_{\rm max}^2/\omega^2\approx
10^{10}$. So the dominant term in Eq.~(\ref{Inc}) is the first one. To have
an  oscillating solution ($I>0$) it is sufficient to have a very slow
latitudinal differential rotation, $\beta\geq 10^{-8}\approx O(\omega\,[{\rm
s}^{-1}])$. 

For waves running in opposite direction to rotation in the azimuth 
($\beta\bar{k}_x>0$), the cavities (trapped wave area) can form
between the bottom of the convective zone and almost the center of the Sun 
as well as in the outer part of the Sun where $N^2>0$. 

Waves propagating parallel to rotation may
be trapped only in the convective zone ($N^2<0$). To the outer and inner sides
from the convective zone the amplitude of these waves decrease exponentially.      

For the incompressible case it is easy to solve the eigenvalue problem of the
cavity oscillations, because Eq.~(\ref{Inc}) has no singularity. Such a task
has been solved by Oraevsky \& Dzhalilov (1997). However, in the nonadiabatic
case we cannot apply
the limit $c^2\to\infty$.
Therefore we have to investigate the more complicated compressible case.  

To investigate the function $I(z)$ given by Eq.~(\ref{I}) in a 
{\it{compressible}} plasma we need the orders of the quantitities entering
the function $I(z)$. To estimate these values let us consider a linear profile
of temperature, $T_0\approx T_{\rm c}(1-\beta_T z)$, where the gradient
$\beta_T=1-T_{\rm eff}/T_{\rm c}\approx 1$, $T_{\rm eff}$ is the effective
temperature. Then we
have a limit for the parameter $\kap_T$ from the center ($z=0$) up to the
surface ($z=1$): $1\le -\kap_T\le 10^3$. The other parameters have the same
order, $\kap_{p,\rho}\approx O(\kap_T)$. Then 
we get also: $\kap'_T\approx -\kap^2_T, \,\, \kap'_p\approx\kap_p\kap_T, \,\,
\kap'_{\rho}\approx\kap_T(\kap_p+\kap_T)$. Now we can estimate the sign
behavior of $I(z)$. We shall consider more characteristic places of the Sun.
In the following the condition $\beta\bar{k}_x>0$ will be supposed.

{\it{At the center}}, where $z\to 0,\,g\to 0,\,N^2\to 0,\,\kap_p\to 0,\,
c\approx 350$\,km/s, if $\Lambda>1$ we have
\[
I\approx -\beta\bar{k}_x\frac{\omega}{2\tilde{\Omega}}-
\frac{3}{2}\kap_{\Omega}-\frac{1}{4}\left(\frac{1}{2}\kap_T-
\kap_{\Omega}\right)^2-\frac{1}{2}\kap_T^2 < 0 .
\] 

In the {\it{middle part}}  of the Sun (between the edge of the core and the
bottom of the convective zone), where $N^2\approx N^2_{\rm max}$ and 
$\kap_{\rho}\approx {\rm const}\approx -10$, the dominant term in the
function $I(z)$ is the first term of function $\psi_0$. Hence we have $I>0$
such as in the incompressible case .

The area {\it{around the convective zone}} is more complicated.
The function $I(z)$ has singularities at the points where $d=0$ and $a=0$; 
$d$ is connected with the Brunt-V\"ais\"al\"a frequency and
$a$ is defined mainly by the rotation gradient, Eqs. (37').
The two points of $d=0$ correspond to the bottom and the upper boundary
of the convective zone. The function $a(z)$ is defined mainly by two terms
\be
\label{a}
a\approx -\Lambda -\frac{\gamma-1}{\gamma}\frac{d'}{d^2} .
\ee
It is clear that as $d'<0$ at the lower boundary of the convective zone, the
two zeros of $a(z)$ are located around this boundary where $d=0$.
We can easily conclude that
\[
\lim_{a\rightarrow 0} I(z)\approx -
\frac{3}{4}\left(\frac{a'}{a}\right)^2\to -\infty,
\]
\be
\label{lim}
\lim_{d\rightarrow 0} I(z)\approx 2\left(\frac{d'}{d}\right)^2\to +\infty .
\ee

In the convective zone ($N^2<0$) we have $I<0$. At the surface of the Sun
we have again $N^2>\omega^2$ and hence $I>0$.

\begin{figure}
\vspace{7cm}
\includegraphics{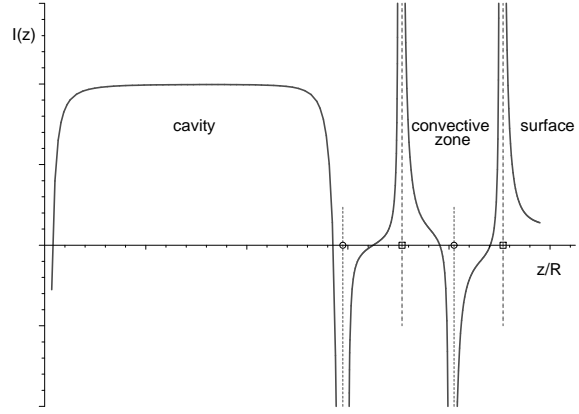}
\caption{Scheme of the dependence of the wave potential $I(z)$ 
defined by Eq.~(\ref{I}) on the distance normalized to the solar radius. 
Zeros of this function $I(z)=0$ (turning points) divide the wave zone into 
transparent (cavity) and opaque (tunnel) parts. In the upper part of the Sun 
there are singular points of the function $I(z)$ which are located 
between the turning points: the circles correspond to $a(z)=0$ and the boxes 
correspond to $N^2(z)=0$ (boundaries of the convective zone). The narrow area
at the bottom of the convective zone between circles and boxes is the
tachocline. The main internal cavity comprises the whole radiative
interior as the convective zone becomes opaque for the waves.  }
\end{figure}

Thus if we approach the convective zone from below there exists a sequence 
of special points: four times the function $I(z)$ crosses zero (turning
points), and between these zeros the singular points are placed (see Fig.1).
The turning points are determined practically by the condition of 
$N^2(z)-\omega^2 =0$ and the singularities by the conditions $a=0$ (circles in
Fig.1) and $N^2=0$ (boxes in Fig.1). Due to the very low frequency and the
sharp decrease of $N^2(z)$ the turning point of the main inner cavity $z_t$ is
very close to the first singular point where $a=0$.

In this way for waves with $k_x>0$ a main large internal cavity is placed
practically between the center and the bottom of the tachocline. The solar
atmosphere is a wave-propagating zone. Between the inner cavity and the solar 
surface a dark convective tunnel is placed; a very narrow wave-trapping zone
around the bottom of convective zone is also possible. It is clear that a 
tunneling of waves across the magnetized and turbulent convective area to the
surface is probably possible. For waves with $k_x<0$ the convective zone
becomes a cavity.
In this case the waves cannot be propagating at the solar surface.  

Not all singularities of the wave potential $I(z)$ are singular levels of the
physical variables. Around the point $z_d$ where
$d(z_d)=0$ we can write $d\approx d'(z-z_d)$. Then we have the equation
$x^2Y'' +2 Y=0$, where $x=z-z_d$. The solutions of this equation are
$Y_{1,2}=\sqrt{x}x^{\pm i\sqrt{7}/2}$. As $a\sim d^{-2}$ we get from
Eqs.~(\ref{ap'} + \ref{Y}) that $p'/p_0\sim \Theta\sim Y_{1,2}\to 0$ if
$z\to z_d$. It means, that the boundaries of the convective zone are not
singular levels for the initial physical variables.

Another situation exists at the point $z=z_a$ where $a=0$. If we denote now
$x$ as $x=z-z_a$, our Eq.~(\ref{Y}) around the point $x\approx 0$ is
$x^2Y''-Y3/4=0$, the solutions of which are $Y_1=x^{3/2},\,Y_2=x^{-1/2}$.
Hence, for these solutions
we have $\Theta_1\approx x^2,\,\Theta_2\approx$ const. Then we get from
Eq.~(\ref{ap'}) that the second solution in $p'/p_0$ diverges at $a=0$.

There exist methods to construct asymptotic solutions of differential
equations of second order with a singular turning point. However, we are now
returning to our fourth order Eq.~(\ref{weq}) for two reasons: our intent is
to consider the instability problem of the eigenmodes, and consequently in the
complex $\omega$ plane the singularity at $a(z,\omega)=0$ is removed from
the real  $z-$axis.


\section{Asymptotic solutions}


The existence of the small parameter $\en$ in Eq.~(\ref{weq}) allows us to
apply asymptotic methods to solve this equation. Here we shall construct
the inner cavity solutions only. As it has been discussed above the 
coefficients of Eq.~(\ref{weq}) vary over a wide range. Very crudely, we have
the following estimates from the center of the Sun to the bottom of the
convective zone : $\en\approx O(10^{-8} - 10^{-5}),\,\kap_T\approx O(1 -
10^3),\,\varphi_3\approx O(\kap_T),\,\varphi_2\approx 
O(\kap_T^2),\,\varphi_1\approx O(\kap_T^3),\,\varphi_0\approx 
O(\kap_T^4),\,\psi_1\approx O(\kap_T),$ and $\psi_0\approx O(\kap_T^2)$. 
Assuming that $\en\kap_T^2\approx O(10^{-8} - 10)$ we can separate the
variable part of $\en$ as
$\en=\tilde{\en}T_c/T_0(z)$ and rewrite Eq.~(\ref{weq}) in a convenient form:
\begin{eqnarray}
\tilde{\en}(\Theta''''&+&\Phi_3\Theta''')+\Phi_2\Theta''+\Phi_1\Theta'+
\Phi_0\Theta =0, \label{wq1} \label{weq1}\\
\Phi_0&=&\frac{T_0}{T_c}(\psi_0+\en\varphi_0),\,\,
\Phi_1=\frac{T_0}{T_c}(\psi_1+\en\varphi_1),\,\, \nonumber\\
\Phi_2&=&\frac{T_0}{T_c}(1 + \en\varphi_2),\,\, \Phi_3=\varphi_3.\nonumber
\end{eqnarray}
The general solution of this equation describes the slow and fast oscillating
motions coupled with each other (quasi-adiabatic and dissipative modes).
Every of these solutions is searched in a different way.


\subsection{Dissipative modes}


The solutions corresponding to the dissipative modes are searched for in the
form:
\be
\Theta=B_0(z)\left(1+\tilde{\en}^{1/2}B_1+
\tilde{\en}B_2+\cdots\right) e^{\tilde{\en}^{-1/2}\int^zS(z)\,dz}
,\label{dm0} \ee
where $S(z)$ and $B_{0,1,2,\ldots}(z)$ are unknown regular
functions. Inserting the formal solution (\ref{dm0}) into Eq.~(\ref{weq1})
we may  find all the unknown functions by the usual methods. In particular
we have
\begin{eqnarray}
S^2&=&-\Phi_2, \ \ SB_0=\Phi_2^{-3/4}e^{\frac 12\int^zF_0\,dz},\\
B_1&=&\frac 12\int^z\frac{F_1}{SB_0\Phi_2}\,dz, \\
F_0&=&\frac{\Phi_1}{\Phi_2}-\Phi_3 =\frac{\psi_1+
\en\varphi_1}{1+\en\varphi_2} -\varphi_3,\\
F_1&=&S(2SB_0''+3S'B_0'+S''B_0)+3[S(SB_0)']'+ \nonumber\\
& & + 3\Phi_3S(SB_0)'+\Phi_1B_0'+ \Phi_0B_0 .
\end{eqnarray}
In the derivation of these functions we have used the obvious condition
$\tilde{\en}^{-1/2}SB_0\gg B_0'$. It may easily be shown that at the boundaries
of the convective zone where $d=0$ the solution (\ref{dm0}) is limited. 
To solve the eigenvalue problem we can restrict ourselves to the main branch
of the asymptotic solutions of (\ref{dm0}):
\be
\label{dm1}
\Theta\approx \left(\frac{T_c}{T_0}\frac{1}{1+\en\varphi_2}\right)^{5/4}
e^{\frac 12\int^zF_0dz}e^{\pm i\int^z(\frac{1}{\en}+\varphi_2)^{1/2}dz} .
\ee 
In order to get the final form of this solution let us consider the extreme
case $\en\to 0$. Then we have
\be
\label{dmA}
\Theta\approx \frac{\sqrt{\rho_0}}{T_0}\exp\left(\pm i\int^z\frac{dz}
{\sqrt{\en}}\right) ,
\ee
where $1/\en=i/|\en|$ and $1/\sqrt{\en}=\pm(1+i)/\sqrt{2|\en|}$. It is clear
that the turning points for these waves do not exist in the usual sence. 
It means, that these waves must propagate from a source down and up by
decreasing their amplitudes. Hence only the `transition' turning point for the
dissipative modes can appear. The details of this question have been discussed
by us in a recent paper (Dzhalilov et al. 2000) for the $p$-modes leakage
problem. The place of the location of the
sources of these waves can be taken in those places where the adiabatic
approximation fails. Around the upper turning point $z=z_t$ the waves are rather
nonadiabatic, where their vertical wavelengh is large.  

Thus we have a radiation boundary condition for the dissipative modes: from the
point $z=z_t$ (turning point of adiabatic waves) the dissipative waves
are radiated. In all directions the amplitudes of these modes
must decrease, and while setting $\en\equiv 0$ these modes must disappear.
The following solution obeys these conditions: 
\begin{eqnarray}
\Theta&=&\Theta_{D}\approx \frac{T_0^{1/4}}{(1+\en\varphi_2)^{5/4}}
e^{J-J_0},\label{dm2}\\
J_0(z)&=&\frac 12\int_z^{z_t}\en\,G(z)\,dz, \ \  
G(z)=\frac{\varphi_1-\psi_1\varphi_2}{1+\en\varphi_2}, \label{J0} \\ 
J(z)&=&i\int^z_{z_t}\left(\frac{1}{\en}+\varphi_2\right)^{1/2}dz,\,\,
\mbox{if}\,\, z\ge z_t,\nonumber \\
J(z)&=&i\int^{z_t}_z\left(\frac{1}{\en}+\varphi_2\right)^{1/2}dz,\,\,\mbox{if}\,\, z\le
z_t.
\end{eqnarray}
Here the branch with $\sqrt{2i}=+(1+i)$ is choosen, Re$(J)<0$, and $T_0(z)$ is
normalized to $T_{\rm c}$. 


\subsection{Quasi-adiabatic modes} 


Now we will search for `slow' quasi-adiabatic wave solutions in the form
\be
\Theta=\sum\limits_{n=0}^{\infty}\tilde{\en}^nY_n(z) ,\label{Ad1}
\ee 
where the functions $Y_n(z)$ are supposed to be smooth. Inserting this
formal solution into Eq.~(\ref{weq1}) we receive the reccurent
differential relations between the $Y_n$ functions:
\begin{eqnarray}
\Phi_2Y_0''+\Phi_1Y_0'+\Phi_0Y_0&=&0, \label{Aeq0} \\
\Phi_2Y_n''+\Phi_1Y_n'+\Phi_0Y_n&=&-\left(\der{}{z}+\varphi_3\right)\frac{d^3}{dz^3}Y_{n-1},
\end{eqnarray}
where $n=1,2,\cdots$. For our aim we are interested in the main value
of the solution (\ref{Ad1}) only. Therefore for small $\en$ Eq.~(\ref{Aeq0})
describes the main properties of the quasi-adiabatic waves.
To obtain the standard form of this equation we introduce a new variable
\be
Y_0=\exp\left(-\frac 12\int^z\frac{\Phi_1}{\Phi_2}dz\right)\,W=
T_0^{3/4}\sqrt{a}\,|d|\,e^{J_0}\,W , \label{Y0}
\ee 
where $J_0(z)$ is defined by Eq.~(\ref{J0}). Then we have
\be
W''+(\lambda^2\wp +\tilde{\varphi})W=0 ,  \label{Aeq1}
\ee
where $\lambda^2=c_0^2\Lambda/c^2$, $\tilde{\omega}_0=\omega R_{\odot}/c_0$,
$c_0$ is given for the temperature in the solar center, and
\begin{eqnarray}
\wp(z)&=&\frac{\Lambda-\alpha_0}{1+\en\varphi_2}\frac{\tilde{\omega}_0^2}{\Lambda}
\left(\frac{N^2}{\omega^2}-1+\en\tilde{k^2_{\perp}}\right) , \\
\alpha_0&=&1+\left(\kap_{\rho}+\der{}{z}\right)\frac{q}{d}+
\left(\tilde{\omega} \frac{q}{d}\right)^2 \approx -q\frac{d'}{d^2} , \\
\tilde{\varphi}(z)&=&\frac{b_1(a_1+\kap_T+b_1'/b_1)+
\en\varphi_0}{1+\en\varphi_2}- \nonumber\\
& & -\frac 14(\psi_1+\en G)^2 -\frac 12\der{}{z}(\psi_1+\en G) .
\end{eqnarray}
Here $\lambda=$const is a large spectral parameter, and $\tilde{\varphi}(z)$ is
a smooth function. The asymptotic theory for differential equations of second
order such as Eq.~(\ref{Aeq1}) is well developed for real $\wp$. In our case
the function $\wp$ is complex-valued with a complex spectral parameter $\omega$. In 
such a situation the direct application of the standard asymptotic theory to
Eq.~(\ref{Aeq1}) is impossible. For the complex-valued $\wp$ the turning
points of Eq.~(\ref{Aeq1}) are shifted from those of the adiabatic
Eq.~(\ref{Y}). Such small shifts are very important for the solution of the
frequency discrepancy problem of helioseismology (Dzhalilov et al. 2000). For
our task here the small
shifts of the turning points are unimportant. Thus we take for
Eq.~(\ref{Aeq1}) the same turning points which we have discussed in the
section of adiabatic waves. For the solutions we apply the same method as
earlier for the $p$-modes (Dzhalilov et al. 2000). We get the solutions
\begin{eqnarray}
W&=&\left(\frac{\xi}{\wp}\right)^{1/4}\left[C_1A_i(-\xi)+C_2B_i(-\xi)\right],\,\,z\le
z_t , \label{Ai0}\\
\xi&=&\left(\frac 32\eta\right)^{2/3},\,\,
\eta=\lambda\int_z^{z_t}\sqrt{\wp}\,dz ,\,\, 
\xi'(z)=-\lambda\sqrt{\frac{\wp}{\xi}}, 
\end{eqnarray}
where $\xi'(z_t)\ne 0 $ and
\begin{eqnarray}
W&=&\left(-\frac{\xi}{\wp}\right)^{1/4}\left[C_1A_i(\xi)+C_2B_i(\xi)\right],\,\,z\ge z_t ,\label{Bi0}\\
\xi&=&\left(\frac 32\eta\right)^{2/3},\,\,\eta=\lambda\int^z_{z_t}\sqrt{-\wp}\,dz,\,\,
\xi'(z)=-\lambda\sqrt{-\frac{\wp}{\xi}} . \nonumber
\end{eqnarray}
Here the regular functions $A_i(\xi)$ and $B_i(\xi)$ are the Airy functions of
first and second kind and $C_{1,2}=$const. In the derivation of the solutions
(\ref{Ai0} + \ref{Bi0}) the branch $-1=\exp(i\pi)$ has been taken.

Now we have to determine from the boundary conditions one of the unknown
constants $C_{1,2}$. Here we assume the next simplification. In the real solar
situation we should include the tunneling of waves through the opaque
convective zone to the transparent solar surface. However, in this work we do
not complicate the situation by including this important effect. Here we are
interested in the eigenoscillation spectrum of the main interior cavity. So our
solutions must be finite in the whole domain of integration: $0\le z \le 
\infty$. In the limit of $z\to\infty$ we have the asymptotics
\be
A_i(\xi)\approx\frac{1}{2\sqrt{\pi}}\xi^{-1/4}e^{-\eta} , \ \ 
B_i(\xi)\approx\frac{1}{\sqrt{\pi}}\xi^{-1/4}e^{\eta} ,
\ee
where $\eta$ is complex and Re$(\eta)\to\infty$. From here we have the
condition $C_2=0$, and the solutions (\ref{Ai0} + \ref{Bi0}) are expressed
only by the function $A_i(\xi)$. For small $\en$ we have the cavity solution
for the quasi-adiabatic waves from Eqs.~(\ref{Ad1}), (\ref{Y0}),
and (\ref{Ai0}): 
\be
\Theta=\Theta_{NA}\approx T_0^{3/4}\sqrt{a}\,|d|\,e^{J_0}
\left(\frac{\xi}{\wp}\right)^{1/4} A_i(-\xi) , \label{sna}
\ee
where $T_0(z)$ is normalized to $T_{\rm c}$. This solution is limited at the
boundaries of the convective zone ($d=0$). Now we can define the general
solution of our general Eq.~(\ref{weq}) as a superposition of the
solutions for the non-adiabatic (Eq.~(\ref{sna})) and the dissipative
(Eq.~(\ref{dm2})) cases: 
\be
\Theta=C_1\Theta_{NA} + C_2\Theta_D , \label{sum}
\ee
where the new arbitrary constants $C_1$ and $C_2$ must be determined from
the boundary conditions.


   \section{Boundary value problem}


In this section we impose physically reasonable boundary conditions
to the general solution of (\ref{sum}) to determine the spectrum of the
eigenoscillations of the inner cavity.


\subsection{Solar center}


At the solar center ($z=0$) where $g\to 0, \kap_p\to 0$, and
$\kap_{T,\rho}=O(1)$
we apply a rigid boundary condition: $v_z(0)=0$. As the function $\Theta(z)$ is 
finite at the center Eq.~(\ref{vz}) reads as $\dertext{(p'/p_0)}{z}=0$. 
Using Eqs.~(\ref{p'} + \ref{f}) with dimensionless parameters we have
\be
\der{}{z}\Theta\left(1+\en\frac{\Theta''}{\Theta}\right)=0 . \label{can0}
\ee
It is easy to show that for both solutions the conditions 
$\Theta_{NA}''/\Theta_{NA}\approx $const,\, $\Theta_D''/\Theta_D\approx$ const
are obeyed. Then the condition (\ref{can0}) is changed to
\be
\left.\der{\Theta}{z}\right|_{z=0}=0. \label{can1}
\ee  
Now we insert the general solution (\ref{sum}) into this condition and receive
the ratio between the unknown coefficients:
\be
\frac{C_2}{C_1}=-\frac{\Theta'_{NA}(0)}{\Theta'_D(0)}. \label{C21-1}
\ee
To obtain the dispersion relation we need one more boundary condition.


\subsection{Bottom of the convective zone}


Let us consider the location of the upper turning point ($z=z_t$): there the
equation $N^2\approx\omega^2$ is fulfilled, from which we may define the
parameter $d$, $d\approx O(R_{\odot}\omega^2/g)\approx 10^{-9}\approx 0$.
Here we can  apply a free boundary condition, because at the bottom of the
convective zone the Brunt-V\"ais\"al\"a frequency and the rotation rate
$\Omega(z,\theta)$ change very sharp. (A more realistic approach would be to
include the tunneling of the waves.) At the free surface the Lagrangian
pressure must be constant, $dp/dt=0$, where $p=p_0 +p'$ is the total
pressure. From here we have 
\be 
v_z=-\frac{i\omega c^2}{g}\frac{p'}{p_0}\bigg|_z=z_t .
\label{cnd2-0} 
\ee
From Eqs.~(\ref{ap'} + \ref{p'}) we define the next formulae written
in dimensionless parameters
\begin{eqnarray}
a\kappa\frac{p'}{p_0}&=&\Theta+\frac{q\tilde{\omega}^2}{d^2}
\left(1-\frac{N^2}{\omega^2}+\en\frac{d^2}{dz^2}\right)\Theta+ \nonumber \\
 &+&\frac{1}{d}\left(\kap_{\rho}-\frac52\kap_T-\frac{d'}{d}+\der{}{z}\right)
\left(1+\en\frac{d^2}{dz^2}\right)\Theta , \label{kp'}\\
\frac{v_z}{c}&=&\frac{i\tilde{\omega}}{\kappa d}\left[\left(1+
\en\frac{d^2}{dz^2}\right) \Theta-\kappa q\frac{p'}{p_0}\right] . \label{vz-c}
\end{eqnarray}
Having in mind that at $z\approx z_t$ we have $a\approx -qd'/d^2$ and $d'<0$;
neglecting small values we get from Eqs.~(\ref{cnd2-0}) + (\ref{vz-c})
\be
\left(1+\en\frac{d^2}{dz^2}\right)\Theta=\kappa q\frac{p'}{p_0} .
\ee
Now excluding $p'/p_0$ from here and from Eq.~(\ref{kp'}) we get the second 
boundary condition
\be
\left.\der{\Theta}{z}\right|_{z=z_t} \approx 0 . \label{can2}
\ee
Inserting the general solution (\ref{sum}) into this condition we get
\be
\frac{C_2}{C_1}=-\frac{\Theta'_{NA}(z_t)}{\Theta'_D(z_t)}. \label{C21-2}
\ee


\subsection{Dispersion relation}


From Eqs.~(\ref{C21-1} + \ref{C21-2}) we get the general dispersion
relation
\be
\label{dis0}
\Theta'_{NA}(0) =\Theta'_D(0)\frac{\Theta'_{NA}(z_t)}{\Theta'_D(z_t)} .
\ee
To simplify this equation the asymptotic expansions of the Airy functions
can be used. At the center of the Sun the arguments of the Airy functions
become large: $|\xi|>>1$. Then the main values of the asymptotic expansions
yield (Eqs. 68 + 53)
\begin{eqnarray}
\Theta'_{NA}(0)&\simeq &
-\frac{i\lambda^2\wp_c^{1/4}}{\sqrt{\pi}}\cos\left(\eta_0+\frac{\pi}{4}\right)
, \label{Ai-0} \\
\Theta'_D(0)&\simeq &
-\frac{i}{\sqrt{\tilde{\en}}}\exp{\left(\frac{i}{\sqrt{\tilde{\en}}}
\int_0^{z_t}\frac{T_0}{T_c}dz\right)} , \label{D0}
\end{eqnarray}
where $\eta_0=\eta(0)$ (Eq. 65) and $\wp_c=\wp(0)\approx -\tilde{\omega}_0^2$
(Eq. 61). At the turning point $z=z_t$ we have
\begin{eqnarray}
\Theta'_{NA}(z_t)&\simeq &
\left(\frac{T_t}{T_c}\right)^{3/4}(-d'q)^{1/2}(\wp'(z_t))^{1/6}
\lambda^{5/6}A_i'(0) , \label{Ait} \\
\Theta'_D(z_t)&\simeq & -\left(\frac{T_t}{T_c}\right)^{3/4}\frac{i}{\sqrt{\tilde{\en}}} ,
\label{Dt}
\end{eqnarray}
where $T_t=T_0(z_t)$. Now setting Eqs.~(\ref{Ai-0}) -(\ref{Dt}) into
(\ref{dis0}) we get
\begin{eqnarray}
\cos(\eta_0+\frac{\pi}{4})&\simeq &
i\sqrt{\pi}A_i'(0)\frac{\wp'(z_t)(-d'q)^{1/2}} {\lambda^{7/6}\wp_c^{1/4}}
\times \nonumber \\
 &&\times \exp{\left[\frac{i-1}{\sqrt{2|\tilde{\en}|}}
\int_0^{z_t}\left(\frac{T_0}{T_c}\right)^{1/2}dz\right]} . \label{dis1} 
\end{eqnarray}
The right-hand side of this equation is negligible small as
$Ai'(0)=-0.259$, $\lambda\gg1$ (Eq. 60), $q\approx(\gamma-1)/\gamma$
(Eq. 37'), $d'(z_t)\approx 16$, and mainly  $\tilde{\en}\approx 10^{-8}$ (Eq.
38). Thus we have finally using Eq. (65)
\be 
\lambda\int_0^{z_t}\sqrt{\wp(z)}dz=\pi(n-\frac
34) . \label{dis2} 
\ee 
We may rewrite this dispersion relation for the
$R$-modes in a visually more convenient form
\begin{eqnarray}
\frac{\omega}{2\Omega_{\odot}}&=&\frac{\beta\bar{k}_x}
{\bar{k}^2_{\perp}+\sigma_n^2} ,\label{dis3}\\
{\it{I}}(\omega,k_x)&=&\int_0^{z_t}\frac{N}{2\Omega_{\odot}}
\sqrt{A\left[1-\frac{\omega^2}{N^2}\left(1-
\en\tilde{k_{\perp}^2}\right)\right]}\,dz , \nonumber \\
A(z,\omega)&=&\frac{1-\alpha_0/\Lambda}{1+\en\varphi_2},\,\,
\sigma_n= \frac{\pi(n-3/4)}{{\it{I}}} . \nonumber
\end{eqnarray}
Here $n=1,2,\cdots $ are the radial node numbers, for the other
parameters see Eqs. (37' + 62). In the adiabatic limit 
($\en=0$ and then Im$(\omega)=0$) for very small frequencies 
($\omega^2\ll N^2$) and for incompressible motions ($c^2\to\infty$) 
we get the classical dispersion relation for Rossby waves: 
$\omega\sim\beta k_x/(k_{\perp}^2 + k_z^2)$  with $k_z^2=\sigma_n^2$.


\section{Instability of 22-year, 4000-yr, and quasi-biennial oscillations}


Now we can determine the complex eigenfrequencies 
$\omega=\omega_n(\beta,k_x,k_y,\alpha)$ from the integral dispersion relation
Eq.~(\ref{dis3}) for the given $\Omega_{\odot}=2.86\times 10^{-6}$\,s$^{-1}$.
Remind that $\beta$ is the latitudinal gradient of the rotation rate, $\Omega
\approx \tilde{\Omega}(1- \beta y/R_{\odot})$, and $\alpha$ determines the
power exponent of the temperature of the nuclear reactions, $Q\sim \rho^2
T^{\alpha}$ (Eq. 5).  For the $p$-$p$ reactions we will use $\alpha=4$.
The dependence of this equation on the free parameter $k_y$ is a drawback of
this equation. In real situations the waves cannot be progressive across the
shear in $y$-direction. In the general case the complex
$k_y=k_y(k_x,\beta,\omega)$ is a solution of the two-dimensional eigenvalue
problem. To simplify our task for a very small latitudinal gradient of
rotation $\beta$ we have introduced the free parameter $k_y$. To simulate the
decay of the wave amplitudes towards the pole, in $y$ direction, we shall
consider only the case with real $k_y^2<0$ . A negative spectral parameter
$k_y^2$ could also be confirmed by the tidal equation of Laplace (e.g. Lee \&
Saio 1997) for very low frequencies. This equation includes the influence of
sphericity on the angular dependence of the eigenfunctions, if simple rigid
rotation is considered in the frame of the traditional geophysical
approximation (which fails near the equator). 

Ando (1985) has derived a local dispersion relation for waves around the
stellar equator. Solving his Eq.~(15) with respect to $k$ (that is our
$k_{y}$) for low frequencies $(\omega/2\Omega \ll 1)$ we get two solutions:
$k^2 = -m^2$ and $k^2 = -m^2N^2/4\Omega^2$, where $m$ is the azimuthal wave
number. As $N^2/4\Omega^2 \gg 1$, the second solution is strongly damped in
$y-$direction. Our case corresponds to the first solution, $k^2 = -m^2$.

From geophysical applications (Pedlosky 1982; Gill 1982) we infer that
$k_y^2(k_x)\sim -k_x^2$ is obeyed for the ageostrophic wave propagation
across the shear flow. Really, it is seen from Fig.2, that the mode
separation with respect to $n$ is essential for the solar situation if the
condition $k_y^2\approx -k_x^2$ is fulfilled. In Fig.2 the Re$(\omega(k_y))$
dependence is shown for the case $\bar{k}_x=1$ and $\beta=10^{-6}$ for
example. Here, a very small imaginary part of the frequency is supposed,
$\eta\ll 1$, where \be \omega=\omega_r(1+i\eta),\,\,\, \eta={\rm
Im}(\omega)/{\rm Re}(\omega). \ee  In the case $k_y^2/k_x^2\ne -1$ we have a
continuous spectrum which is physically not interesting. So during this work
we have excluded the parameter $k_y^2$ from our dispersion relation
(\ref{dis3}) using $k_y^2= -k_x^2$. As a result the frequencies depend only
on the product $\beta k_x$; the dependence on $\alpha$ is very weak.

\begin{figure}
\vspace{75mm}
\includegraphics{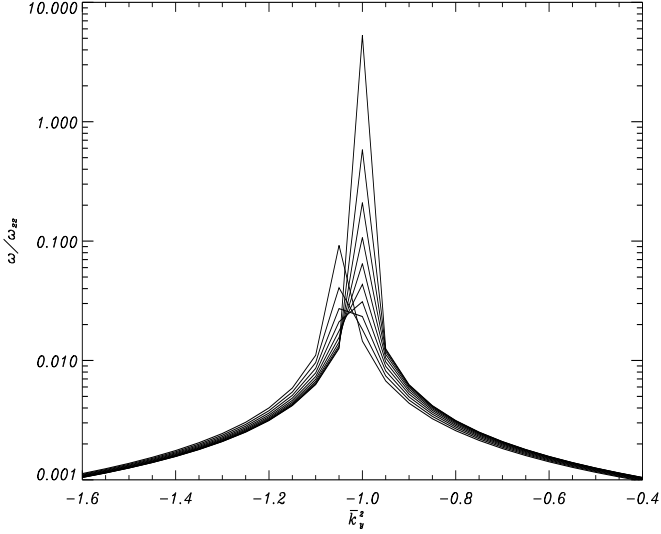}
\caption{Mode separation region versus $\bar{k}_y=k_yR_{\odot}$.
On the vertical axis the absolute values of the frequencies normalized to 
$\omega_{22}$ are shown. The maxima from top to down in the curves 
correspond to n=1, 3, 5, ...\,, 19. Calculations are done for the case
$\bar{k}_x=1, \, \beta=10^{-6},\,\eta=10^{-5}$. This figure does not
yet show the eigenfrequencies, but the domain of $k_y^2$ where the $n$-dependence
of the frequencies is obvious. This domain is $k_y^2\approx -k_x^2$. }
\end{figure}

\begin{figure}
\vspace{75mm}
\includegraphics{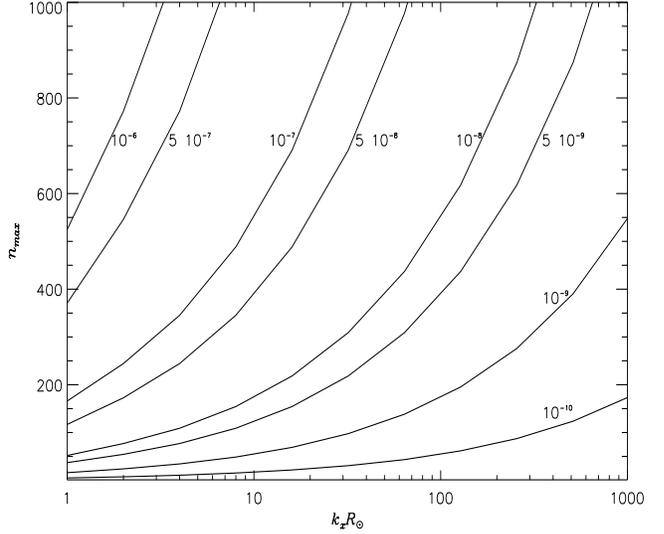}
\caption{ The  harmonic numbers ($n_{\rm max}$) limited by the nonadiabatic
effects versus $k_x$ and $\beta$. The quantities on the curves
correspond to the gradients of the rotation rate $\beta$. Such a limit
for high $n$ does not exist for the adiabatic waves.}
\end{figure}

\begin{figure}
\vspace{149mm}
\includegraphics{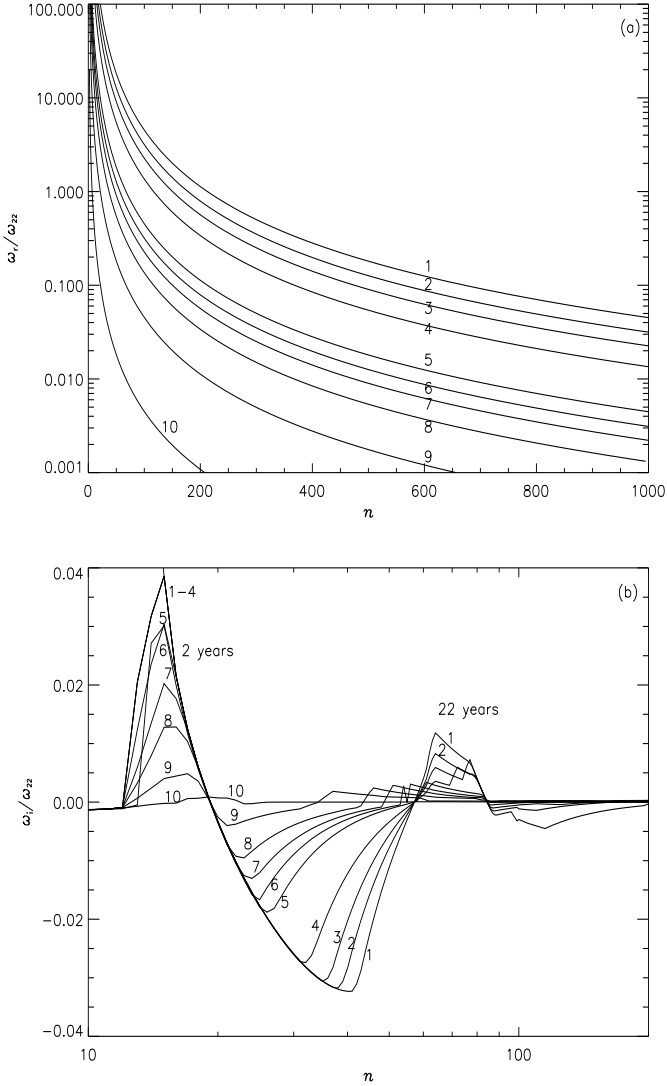}
\caption{Real ($\omega_r$) and imaginary ($\omega_i$) parts of the
eigenfrequencies ((a) and (b), respectively) normalized to $\omega_{22}$
 versus the radial harmonic
number $n$. The numbers on the curves 1, 2, ... , 10 correspond to different
values of the latitudinal gradient of the rotation rate 
$\beta_{1-10} = 10^{-4}, 7\times 10^{-5}, 5\times 10^{-5}, 3\times 10^{-5},
10^{-5}, 7\times 10^{-6}$, $5\times 10^{-6}, 3\times 10^{-6}, 10^{-6},
10^{-7}$. Positive $\omega_i$ corresponds to the growth rate of the waves due
to the $\en$-mechanism. For negative $\omega_i$ the waves become damped in
time due to radiative losses.  }
\end{figure}

\begin{figure}
\vspace{75mm}
\includegraphics{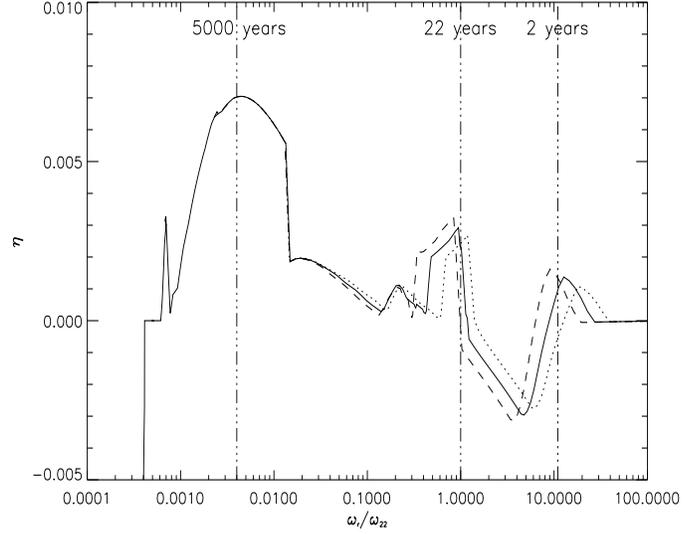}
\caption{Growth rate of the instabilities of the eigenmodes of the
differentially rotating Sun. Three modes with periods of $\approx$\,1--3\,yr,
18--30\,yr (these ranges depend on $\beta\bar{k}_x$) with a small additional
peak at 100 yr, and 1500--20\,000\,yr (independent of $\beta\bar{k}_x$) 
become maximum unstable ($\eta>0$) for high orders $n$. The solid, dashed,
and dotted lines correspond to the small latitudinal gradients of the
rotation rate of $\beta= 7\times 10^{-6}, 5\times 10^{-6}$, and $10^{-5}$ at
$\bar{k}_x=100$. }
\end{figure}

It follows from Eq.~(\ref{dis3}) that for a given $\beta k_x$ the
frequencies $\omega_n$ decrease with increasing harmonic number $n$
($\omega\sim 1/n^2$) in the adiabatic case ($\en=0$). At very high
$n\to\infty$ we have almost steady motions with a frequency $\omega\to 0$.
Howewer, the situation is changed if dissipation is taken into account,
$\en\ne 0$. In this case small-scale motions are quickly damped, very low
frequency oscillations with high $n$ could not become trapped and cannot
manifest themselves as eigenmodes. Really, as the nonadiabatic parameter
$\en\approx 1/\omega$, at very low frequencies the solutions of the initial
equations must have a dissipative character. An investigation of the initial
equations for a steady flow with $\pdtext{}{t}=0$ is a separate task. Here we
restrict ourselves only to our dispersion relation Eq.~(\ref{dis3}) with
increasing $n$. We can rewrite this equation as $1/n^2\approx \omega/I^2$.
With decreasing $\omega$ the right-hand part $\omega/I^2$ decreases, and then
it becomes independent from $\omega$. It means, that $n$ is limited: $n\le
n_{\rm max}$. Hence, it is a results of the dissipative effects, that only
limited modes become trapped. The value of $n_{\rm max}$ depends on $k_x$ and
$\beta$. In Fig.3 the $n_{\rm max}(k_x,\beta)$ dependence is presented. It
follows from this figure that for a given $k_x$ the number $n_{\rm max}$
increases strongly by increasing the rotation gradient $\beta$. So, if the
solar interior is rotating similar to a solid-body, very long-period
oscillations (almost steady flows) should be suppressed. As the values of
$n_{\rm max}$ are sufficiently high, the accuracy of the asymptotic solutions
is high. 

Now we consider which spectrum of trapped waves with $1\le n\le n_{\rm max}$ is 
possible. To calculate the wave spectrum from Eq.~(\ref{dis3}) we use the
standard model of Stix (private communication; Stix \& Slaley 1990) for the
solar interior. For the special case
$\bar{k}_x=k_xR_{\odot}=100$ (i.e. $\lambda_x\approx 40\,000$\,km) the
$n$-dependence of the real and imaginary parts of the eigenfrequencies are
shown in Figs.4a,b. The calculations were done for different small values of
the rotation rate gradient $\beta$, covering a wide range:
$\beta=10^{-7}$\,--\,$10^{-4}$. This is done because we know from helioseismology
only that the $\beta$ parameter is small, but the exact value is not yet
known. In Figs.4a,b $\omega_r=$Re$(\omega)$ and $\omega_i=$Im$(\omega)$ are
normalized to the cycle frequency of the 22-year oscillations:
$\omega_{22}=0.91\times 10^{-8}$\,s$^{-1}= 2\pi(1.45$\,nHz). As expected the
frequencies decrease with $n$ and increase with $\beta$ or $k_x$. The
imaginary parts $\omega_i$ oscillate around the zero value: if $\omega_i >
0$, the waves are unstable and their amplitudes increase with time; in the
opposite case, if $\omega_i\le 0$, we have stable/damped waves. In Fig.4b we
have two positive maxima: the first corresponds to short-period oscillations
of 1--3 yr (`quasi-biennial modes') and the second one to medium-period
oscillations of 18--30-yr (`22-yr modes'). The position of the
quasi-biennial modes versus $n$ is stable and is $\approx 15$. For the 22-yr
modes $n_{22}$ is slightly increased  with an increase of $\beta$. It is seen
from Fig.4a that $n$ must increase to keep the same frequency with increasing
$\beta$.  For smaller frequencies this shift is stronger. Fig.4b shows that
the instability gets stronger if $\beta$ increases: unstable waves become
more unstable and damped waves are stronger damped. Waves with high $n>200$ 
at very low frequencies also show instability which cannot be shown in Fig.4b
due to the scaling.

Of course, $\beta$ will change with the radius in the real solar radiative
interior. Hence, those places, where $\beta$ becomes relatively large, may
become sources of unstable waves. Our calculations can easily be generalized
for any $k_x$ as Eq.~(\ref{dis3}) is a function of $\beta k_x$ only.   

To characterize the mode instability, the behavior of the parameter $\eta$ is
more important. $\eta$ is the growth rate (increment) of the instability of
the modes if $\eta>0$ and the damping rate (decrement) if $\eta<0$. In Fig.5
we present the $\eta(\omega_r)$ dependence for the whole range of frequencies
($\omega_r< \Omega_{\odot}$) for which our asymptotic theory is valid. We
have three distinguished global maxima for the growth rate $\eta>0$ which
correspond to period ranges of $\approx$\,1--3\,yr, 18--30\,yr with a small
additional peak at 100 yr, and 1500--20\,000\,yr (`4000-yr modes') of the
eigenmodes. These modes have radial node numbers near $n=850, 60$, and 15,
respectively. It was already mentioned, that for the unstable modes the value
of $n$ is slightly changed with a change of $\beta$ . So for the unstable
modes $n$ is high and the asymptotic results are reliable. The growth rates
of the 22-yr oscillations are always greater than those of the quasi-biennial
modes. The characteristic growing time for the 22-year modes is $\approx
1000$\,yr as $\eta\sim 0.003$.

\begin{figure}
\vspace{15cm}
\includegraphics{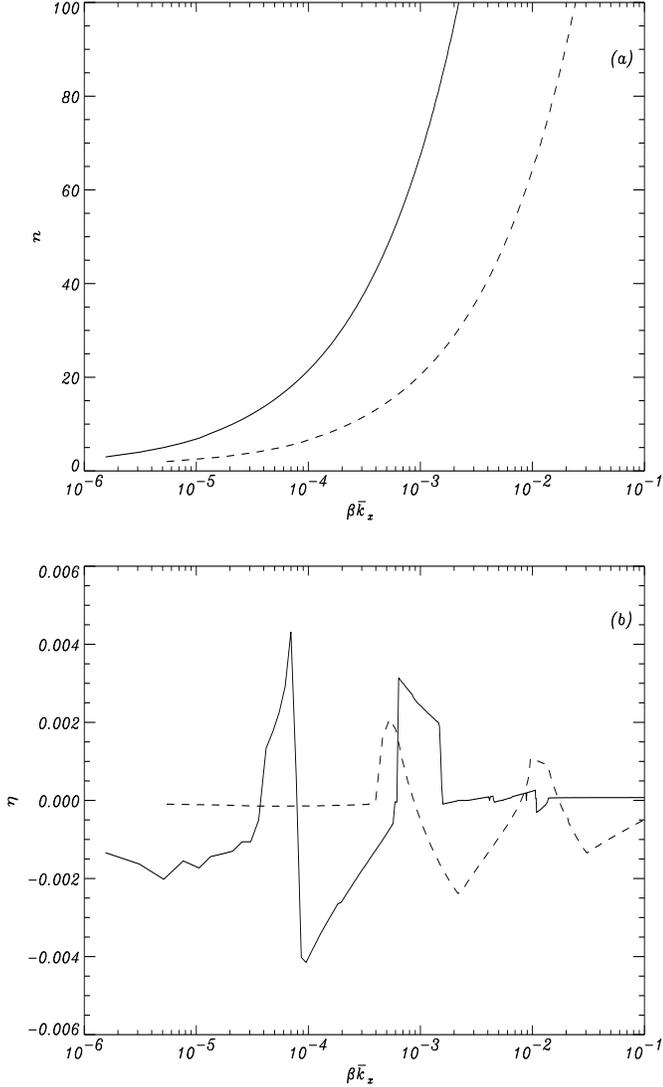}
\caption{The existence area of the 22-yr (solid lines) and 2-yr (dashed
lines) eigenmodes versus the parameters $n, \beta\bar{k}_x$, and
$\eta$. Fig.6a shows that the wavenumber and the rotation gradients 
($\beta\bar{k}_x$) are limited as $n\le n_{\rm max}$. The two narrow maxima of
the growth rate $\eta$ in Fig.6b indicate that for the strongly fixed
values of $\beta\bar{k}_x$ the 22-yr and 2-yr modes become unstable (see text). }
\end{figure}

Now we consider in which range of $n$ and $\beta\bar{k}_x$ the unstable modes
are located. For this aim we fix the frequency as $\omega_r=\omega_{22}=
0.91\times 10^{-8}$\,s$^{-1}$, and for a given $n\le n_{\rm max}$ we find
the complex root of the dispersion relation Eq.~(\ref{dis3}). The parameters 
$\beta\bar{k}_x$ and $\eta$ correspond to this root. The same calculations
were done for the 2-yr oscillations. The results are shown in Fig.6. Here the
solid lines correspond to the 22-yr and the dashed lines to the
quasi-biennial modes. A sharp increase of $n$ to $n_{\rm max}$ with 
$\beta\bar{k}_x$ in Fig.6a   indicates some upper limit for the parameter
$\beta\bar{k}_x$. For the 22-yr and 2-yr modes these limits are approximately
$3\times 10^{-3}$ and $3\times 10^{-2}$, respectively. Fig.6b (where the
increment/decrement is presented) shows for which values of $\beta\bar{k}_x$
and $n$ the 22-yr and 2-yr modes become unstable. Two maxima for each mode
appear in $\eta$. These are for the 22-year modes:\\
 $n=15, \ \ \beta\bar{k}_x =9\times 10^{-5}, \ \ \eta=0.004$ and \\
 $n=55, \ \ \beta\bar{k}_x =7\times 10^{-4}, \ \ \eta=0.003$;\\
 for the 2-yr modes:\\ 
 $n=14, \ \ \beta\bar{k}_x =5\times 10^{-4}, \ \ \eta=0.002$ and\\
 $n=65, \ \ \beta\bar{k}_x =10^{-2}, \hspace{6mm} \ \ \eta=0.001$ .

These estimates could be used to define a possible value of $\beta$ 
inside of the Sun. This will be possible if we can identify these modes
from observations. For instance, at $\bar{k}_x \approx 100$ (sunspot scale)
a gradient of the rotation rate with $\beta=9\times 10^{-7}$ or 
$\beta=7\times10^{-6}$ is needed to excite the 22-yr modes. Here the
possibility of an excitation of both the $n=15$ and the $n=55$ modes is not
excluded if $\beta$ is changed with the radius.
A better way to define $\beta$ would be to identify both the 22-yr and the
quasi-biennial modes, with different $k_x$ at the surface of the Sun.


\section{Conclusions}


In the present paper we have shown that toroidal eddy flows which are
degenerated in a non-rotating fluid can become a reservoir of various branches
of oscillatory modes when the degeneracy is removed by rotation. The mechanism
depends on the condition for the existence and alteration of the relative
vorticity as well as on the stellar rotation rate and its gradient.
Apparently at least for slowly rotating stars ($\Omega<\Omega_g$) the
rotation waves could be divided into two types: $r$-modes with high frequencies
($\omega\le\Omega$) which are independent from the inner structure and mainly
caused by geometrical effects, and the $R$-modes with low frequencies 
($\omega \ll \Omega$) which depend on the inner structure and are considered
in the present paper. This classification is similar to that of $f$- and
$g$- modes or to that of surface and body tube modes of magnetic cylinders.
Note that the properties of Coriolis forces and  ponderomotive forces in MHD
are very similar to each other. Both rotation modes are prototypes of the
geophysical Rossby waves.

We investigated the instability problem of the $R$-modes sustained by a
very small latitudinal gradient of the rotation rate in the solar radiative
interior. Among the eigenoscillations three modes with periods of
$\approx$\,1--3\,yr, 18--30\,yr, and 1500--20\,000\,yr turn out to be
maximum unstable to the $\en$-mechanism. Here the smoothing effect is the
radiative damping. All of these instabilities are in the range of high radial
node numbers $n$ which indicates that the applicability of the asymptotic
solution is satisfied.

The 22-yr modes with a growing time of $\approx1000$~yr are of particular
interest with respect to the solar activity cycle problem. In the simpler
case when adiabatic $R$-modes are considered in an incompressible fluid,
$\sigma_n$ in Eq.~(\ref{dis3}) is independent of the wave number and of the
frequency for very low frequencies. Then in the azimuthal direction the phase
and  group velocities are \be
v_{px}\approx \frac{\beta}{\bar{k}^2_{\perp}+\sigma^2_n}, \ \ 
v_{gx}\approx -\beta\frac{\bar{k}^2_x-\bar{k}^2_y-\sigma^2_n}
{(\bar{k}^2_{\perp}+\sigma^2_n)^2} ,
\ee   
respectively, where the velocities are normalized to
$\Omega_{\odot}R_{\odot}$ $ \approx 1.6$\,km/s.
In our case $\beta>0$, i.e. the angular velocity is decreasing towards the
pole, similar to the behavior at the solar surface. The $x$-axis in our
coordinate system is directed opposite to the direction of rotation,
$v_{px}>0$ and $v_{gx}<0$, moreover, $k_y^2\approx -k_x^2$ and for
finite $n$ we have $\bar{k}_x^2>\sigma_n^2$ (see Fig.2). Then the retrograde
$R$-modes transport energy along the direction of rotation. Our treatment
in a Cartesian coordinate system does not allow to determine the direction of
energy transport by wave packets relative to the equator in the meridian
plane. The estimate of $k_y^2\approx -k_x^2$ is crude, and to determine the
exact dependence on $k_y=k_y(k_x,\omega,\beta)$ the 2D boundary value problem
must be solved.

A nice property of the Rossby waves is that every monochromatic mode is a
solution of the full nonlinear hydrodynamic equations. It means, that
we should expect the development of nonlinear $R$-modes with large amplitudes.
We could also expect that just in this nonlinear regime the toroidal magnetic
flux will be lifted from the upper boundary of the cavity (the tachocline) to
the surface. The energy release of the nonlinear waves could be accomplished
by magnetic reconnection. Here it is possible that toroidal currents are
generated via a twist of toroidal magnetic field lines by the cyclonic flows
of regular $R$-modes with fixed characteristics. Parker (1955) as well as
Steenbeck et al. (1966; see also Krause \& R\"adler 1980) have suggested for
the dynamo process that such a mechanism, the $\alpha-$effect, is working by
turbulent motions under the influence of Coriolis forces.

Our present model points out the possibility of forced oscillations instead
of a self-excited dynamo to solve the solar cycle problem, and this with the
correct period of 22 yr. Similar ideas are due to Tikhomolov (2001) who has
recently suggested a hydrodynamic driving of the 11-yr sunspot cycle.
We expect that in our model --- contrary to classical dynamo models --- a
huge toroidal magnetic field of $\approx 10^5\,G$ will no longer be required
to explain the buoyant rise of magnetic flux tubes appearing at the surface
with small tilt angles and at low latitudes: the external nonzero upflow
produced by the regular vortical $R$-modes could trigger the eruption of
stable magnetic flux tubes stored in the overshoot region. There is still a
smaller peak of the growth rate (Fig.5) at 100 years; such a period is
observed as a modulation of the 11/22 yr cycles.

There is observational evidence for the short-period oscillations as well:
From helioseismic sounding Howe et al. (2000, 2001) have recently discovered
variations of solar rotation with a period of 1.3 yr in the lower convective
zone. Quasi-two year modes are very likely seen regularly in various solar
data (e.g. Waldmeier 1973; Akioka et al. 1987; Rivin \& Obridko 1992). The
existence of two magnetic cycles (the main 22-yr and the quasi-biennial
period) on the Sun has been reported by Benevolenskaya (1996; 1998). So far
the origin of these modes was not yet clear. Terrestrial quasi-biennial
oscillations have been clearly seen in tropical meteorological radiosonde
data, and a possible solar origin by related phenomena in the solar interior,
Rossby waves in particular, has been discussed as well (McIntire 1994). 

The long-period oscillations in the broad range $1.5\times
10^3$\,--\,$2.0\times 10^4$~yr, with a maximum growth rate around 4500~yr,
could be the cause of abrupt changes of the global terrestrial climate in the
past: Dansgaard-Oeschger events, these are abrupt onsets of warm periods
during the last ice age, had mean distances of 4500~yr, but they were
distributed over a larger period range, similar to that in our model, with
shortest distances often around 1500~years (see, e.g., Ganopolski \&
Rahmstorf 2001). These events were caused by changes of the thermohaline
circulation of the ocean, which in its turn were probably triggered by
changes in the solar energy output.


\begin{acknowledgements}
   
Michael Stix kindly provided detailed tables from his internal solar model
calculations. The critical comments and suggestions by Kris Murawski,
Karl-Heinz R\"adler, and the referee J. Andrew Markiel helped to improve
earlier versions of the paper. The authors gratefully acknowledge financial
support of the present work by the German Science Foundation (DFG) under
grant No. 436 RUS 113/560/1-1 and -3, by the German Space Research Center
(DLR) under grant No. 50\,QL\,9601\,9, and by the Russian Foundation for
Basic Research (RFBR) under grant No. 00-02-16271.
\end{acknowledgements}


{}


\begin{thebibliography}{}

\bibitem{} Akioka, M., Kubota, J., Ichimoto, K., et.al., 1987, Solar Phys.
112, 313

\bibitem{} Ando, H., 1985, Publ. Astron. Soc. Jap. 37, 47 

\bibitem{} Baker, N., Kippenhahn, R., 1959, Z. Astrophys. 48, 140

\bibitem{} Benevolenskaya, E.\,E., 1996, Solar Phys. 167, 47

\bibitem{} Benevolenskaya, E.\,E., 1998, ApJ 509, L49 

\bibitem{} Brandenburg, A., 1994, in: Proctor, M.\,R.\,E. \& Gilbert, A.\,D.
(eds.), Lectures on solar and planetary dynamos. Cambridge Univ. Press, p.
117

\bibitem{} Brayn, G.\,H., 1889, Phil. Trans. R. Soc. London A 180, 187

\bibitem{} Caligari, P., Sch\"ussler, M., Moreno-Insertis, F., 1998, ApJ 502,
481

\bibitem{} Cattaneo, F., Hughes, D.\,W., 1996, Phys. Rev.E. 54, R4532

\bibitem{} Chaplin, W.\,J., et.al., 1999, MNRAS 308, 405

\bibitem{} Charbonneau, P., MacGregor, K.\,B., 1993, ApJ 417, 762 

\bibitem{} Charbonneau, P., MacGregor, K.\,B., 1997, ApJ 486, 502

\bibitem{} Cowling, T.\,G., 1953, in: Kuiper, G.\,P. (ed.), The Sun.
Univ.Chicago, p. 565

\bibitem{} Dzhalilov, N.\,S., Staude, J., Arlt, K., 2000, A\&A 361, 1127

\bibitem{} Fan, Y., Fisher, G.\,H., Deluca, E.\,E., 1993, ApJ 405, 390

\bibitem{} Fritts, D.\,C., Vadas, S.\,L., Andreassen, O., 1998, A\&A 333, 343

\bibitem{} Ganopolski, A., Rahmstorf, S., 2001, Nature 409, 153

\bibitem{} Gill, A.\,E., 1982, Atmosphere-Ocean Dynamics. Acad. Press, Univ.
of Cambridge, England

\bibitem{} Gilman, P.\,A., 1969, Solar Phys. 8, 316

\bibitem{} Gilman, P.\,A., 1992, In: Harvey, K.\,L. (ed.), The Solar Cycle.
ASP Conf. Ser. 27, p. 241

\bibitem{} Gough, D.\,O., 1980 in: Hill, H.\,A. \& Dziembowski, W.\,A.
(eds.), Nonradial and Nonlinear Stellar Pulsation. Springer-Verlag, Berlin,
p.273

\bibitem{} Gough, D.\,O., 1985, in: Rolfe, E. \& Battrick, B. (eds.), Future
Missions in Solar, Heliospheric and Space Plasma Physics. ESA SP-235, ESA
Publ. Division, Noordwijk, p.183 
    
\bibitem{} Gough, D., 1997, Nature 388, 324 

\bibitem{} Howe, R., Christensen-Dalsgaard, J., Hill, F., Komm, R.\,W.,
Larsen, R.\,M., Schou, J., Thompson, M.\,J., Toomre, J., 2000, Science 287,
2356

\bibitem{} Howe, R., Christensen-Dalsgaard, J., Hill, F., Komm, R.\,W.,
Larsen, R.\,M., Schou, J., Thompson, M.\,J., Toomre, J., 2001, In: Proc.
SOHO\,10\,/\,GONG\,2000 Workshop `Helio- and Asteroseismology at the Dawn of
the Millenium'; ESA SP-164, p. 45

\bibitem{} Kosovichev, A.\,G., 1996, ApJ 469, L61

\bibitem{} Krause, F., R\"adler, K.-H., 1980, Mean Field
Magneto\-hydro\-dynamics and Dynamo Theory. Pergamon, Oxford

\bibitem{} Kumar, P., Quataert, E.\,J., 1997, ApJ 475, L143

\bibitem{} Kumar, P., Talon, S., Zahn, J.\,P., 1999, ApJ 520, 859

\bibitem{} Lee, U., Saio, H., 1997, ApJ 491, 839 

\bibitem{} Levy, E.\,H., 1992, In: Giampapa, M.\,S. \& Bookbinder, J.\,A.
(eds.), Cool Stars, stellar systems, and the Sun. ASP Conf. Ser. 26, p. 223

\bibitem{} Lockitch, K.\,H., Friedman, J.\,L., 1999, ApJ 521, 764 

\bibitem{} Markiel, J., Thomas, J.\,H., 1999, ApJ 523, 827

\bibitem{} McIntire, M., 1994, in: Nesme-Ribes, E. (ed.), The Solar Engine
and Its Influence on Terrestrial Atmosphere and Climate. NATO ASI Series I
Vol. 25. Springer-Verlag, Berlin-Heidelberg, p. 293

\bibitem{} Mestel, L., Weiss, N.\,O., 1987, MNRAS 226, 123 

\bibitem{}  Moffatt, H.\,K., 1978, Magnetic Field Generation in Electrically
 Conducting Fluids. Cambridge Univ., Cambridge

\bibitem{} Oraevsky, V.\,N., Dzhalilov, N.\,S., 1997, Astron. Rep. 41, 91

\bibitem{} Papaloizou, J., Pringle, J., 1978, MNRAS 182, 423 

\bibitem{} Parker, E.\,N., 1955, ApJ 122, 293
    
\bibitem{} Parker, E.\,N., 1993, ApJ 408, 707

\bibitem{} Patern,o L., Sofia, S., DiMauro, M.\,P., 1996, A\&A 314, 940

\bibitem{} Pedlosky, J., 1982, Geophysical Fluid Dynamics. Springer-Verlag,
New York- Heidelberg-Berlin 

\bibitem{} Provost, J., Berthomieu, G., Rocca, A., 1981, A\&A 94, 126 

\bibitem{} Ringot, O., 1998, A\&A 335, L89

\bibitem{} Rivin, Y.\,R., Obridko, V.\,N., 1992, Astron. Zh. 69, 1083

\bibitem{} R\"udiger, G., Arlt, R., 2000, in: Ferriz-Mas, A. \& Jimenez,
M.\,M. (eds.), Advances in nonlinear dynamos. The Fluid Mechanics of
Astrophysics and Geophysics 8, Chapter 6 (in press)

\bibitem{} Saio, H., 1982, ApJ 256, 717 

\bibitem{} Schatzman ,E., 1993, A\&A 279, 431 

\bibitem{} Schmitt, D., 1993, in: Krause, F., R\"adler, K.-H., \& R\"udiger,
G. (eds.), The Cosmic Dynamo. Kluwer, Dordrecht, p. 1  

\bibitem{} Shore, S.\,N., 1992, An Introduction to Astrophysical
Hydro\-dynamics. Academic Press, Inc. San Diego, Calif., USA

\bibitem{} Smeyers, P., Craeynest, D., Martens, L., 1981, Astrophys. Space
Sci. 78, 483
 
\bibitem{} Spiegel, E.\,A., Zahn, J.\,P., 1992, A\&A 265, 106

\bibitem{} Steenbeck, M., Krause, F., R\"adler, K.-H., 1966, Z. Naturforsch.
21\,a, 369

\bibitem{} Stix, M., 1976, A\&A 47, 243

\bibitem{} Stix, M., 1991, Geophys. Astrophys. Fluid Dyn. 62, 211 

\bibitem{} Stix, M., Skaley, D., 1990, A\&A 232, 234

\bibitem{} Tikhomolov, E., 2001, Solar Phys. 199, 165

\bibitem{} Tikhomolov, E.\,M., Mordvinov, V.\,I., 1996, ApJ 472, 389

\bibitem{} Tomczyk, S., Schou, J., Thompson, M.\,J., 1995, ApJ 448, L57
    
\bibitem{} Unno, W., Osaki, Y., Ando, H., Saio, H., Shibahashi, H., 1989,
Nonradial  Oscillations of Stars. Univ. Tokio Press, Tokyo

\bibitem{} Waldmeier, M., 1973, Solar Phys. 28, 389

\bibitem{} Weiss, N.\,O., 1994, in: Proctor, M.\,R.\,E. \& Gilbert, A.\,D.
(eds.), Lectures on solar and planetary dynamos. Cambridge Univ. Press, p. 59

\bibitem{} Wolff, C.\,L., Bizard J.B., 1986, Solar Phys. 105, 1

\bibitem{} Wolff, C.\,L., 1997, ApJ 486, 1058

\bibitem{} Wolff, C.\,L., 1998, ApJ 502, 961 

\bibitem{} Wolff, C.\,L., 2000, ApJ 531, 591

\bibitem{} Zahn, J.\,P., Talon, S., Matias, J., 1997, A\&A 322, 320

\end{thebibliography}
\end{document}